\PassOptionsToPackage{table,xcdraw}{xcolor}
\documentclass[journal,10pt,compsoc]{IEEEtran}

\usepackage[utf8]{inputenc}
\usepackage[T1]{fontenc}
\usepackage[htt]{hyphenat}  
\usepackage[pdftex]{graphicx}
\usepackage{indentfirst}
\graphicspath{{./images/}{./plots/}{./figures}}
\DeclareGraphicsExtensions{.pdf,.jpeg,.png}
\usepackage{amsmath}
\usepackage{relsize}
\usepackage{mathtools}
\usepackage{amssymb}
\usepackage[d]{esvect}
\usepackage{url}
\usepackage{orcidlink}

\usepackage{caption}
\usepackage[subrefformat=parens]{subcaption}
\usepackage{booktabs}
\usepackage{multirow}
\usepackage{makecell}
\usepackage{tabularx}
\usepackage[table,xcdraw]{xcolor}
\usepackage[detect-weight=true,detect-family=true,binary-units=true,list-units=single,range-units=single]{siunitx}
\usepackage[nolist]{acronym}
\begin{acronym}
    \acro{CPS}{Cyber-Physical System}
    \acro{NCS}{Networked Control System}
    \acroindefinite{NCS}{an}{a}
    \acro{LAN}{Local-Area Network}
    \acro{WLAN}{Wireless Local-Area Network}
    \acro{KPI}{Key Performance Indicator}
    \acro{WPAN}{Wireless Personal Area Network}
    \acro{CSMA/CD}{Carrier-Sense Multiple Access with Collision Detection}
    \acro{CLEAVE}{ControL bEnchmArking serVice on the Edge}
    \acro{OS}{Operating System}
    \acro{UDP}{User Datagram Protocol}
    \acro{TCP}{Transmission Control Protocol}
    \acro{RMS}{Root Mean Square}
    \acro{RTT}{Round-Trip Time}
    \acro{CI}{Confidence Interval}
    \acro{AP}{Access Point}
    \acro{API}{Application Programming Interface}
    \acro{SSF}{Swedish Foundation for Strategic Research}
    \acro{TECoSA}{Trustworthy Edge Computing Systems and Applications}
    \acro{ABC}{Abstract Base Class}
    \acroplural{ABC}[ABCs]{Abstract Base Classes}
    \acro{URL}{Uniform Record Locator}
    \acro{AWS}{Amazon Web Services}
    \acro{EC2}{Elastic Compute 2}
    \acro{AMI}{Amazon Machine Image}
    \acro{SSH}{Secure Shell}
    \acro{IP}{Internet Protocol}
    \acro{VPN}{Virtual Private Network}
    \acro{NAT}{Network Address Translation}
    \acro{NTP}{Network Time Protocol}
    \acroindefinite{NTP}{an}{a}
    \acro{SDR}{Software-Defined Radio}
    \acroindefinite{SDR}{an}{a}
    \acro{VLAN}{Virtual Local-Area Network}
    \acro{APT}{Advaced Packaging Tool}
    \acro{LTE}{Long-Term Evolution}
    \acro{SSID}{Service Set Identifier}
    \acro{EPC}{Evolved Packet Core}
    \acro{UE}{User Equipment}
    \acro{eNodeB}{Evolved Node B}
    \acro{DNS}{Domain Name System}
    \acro{MNIST}{Modified National Institute of Standards and Technology}
    \acro{HTTP}{Hyper-Text Transfer Protocol}
    \acroindefinite{HTTP}{an}{a}
    \acro{YAML}{YAML Ain't Markup Language}
    \acro{O-RAN}{Open Radio Access Network}
    \acro{FOSS}{Free and Open Source Software}
    \acro{COSMOS}{Cloud enhanced Open Software defined MObile wireless testbed for city-Scale deployment}
    \acro{OMF}{ORBIT Management Framework}
    \acro{POWDER}{Platform for Open Wireless Data-driven Experimental Research}
    \acro{COTS}{Commercial Off-The-Shelves}
    \acro{RF}{Radio-Frequency}
    \acro{VM}{Virtual Machine}
    \acro{REST}{REpresentational Sate Transfer}
    \acro{OS}{Operating System}
    \acro{MaaS}{Metal-as-a-Service}
    \acro{OEDL}{\ac{OMF} Experiment Description Language}
    \acro{RSpec}{Resource Specification}
    \acro{LXC}{LinuX Containers}
    \acro{MEC}{Mobile Edge Computing}
    \acro{USRP}{Universal Software Radio Peripheral}
    \acro{OAI}{OpenAirInterface}
    \acro{NR}{New Radio}
    \acro{FPGA}{Field-Programmable Gate Array}
    \acro{WCA}{Wearable Cognitive Assistance}
    \acro{TTF}{Time-to-Feedback}
    \acro{AR}{Augmented Reality}
    \acro{VR}{Virtual Reality}
    \acro{XR}{eXtended Reality}
    \acro{MAR}{Mobile \ac{AR}}
    \acro{GPS}{Global Positioning System}
    \acro{QoE}{Quality of Experience}
    \acro{QoS}{Quality of Service}
    \acro{TTF}{Time-to-Feedback}
    \acro{PCA}{Principal Component Analysis}
    \acro{exGaussian}[Ex-Gaussian]{Exponentially Modified Gaussian}
    \acro{IEEE}{Institute of Electrical and Electronics Engineers}
    \acro{CPU}{Central Processing Unit}
    \acro{N/A}{Not Applicable}
    \acro{CIO}{Confidence Interval}
    \acro{MLE}{Maximum Likelihood Estimation}
    \acro{CDF}{Cumulative Distribution Function}
    \acro{CCDF}{Complementary \ac{CDF}}
    \acro{PDF}{Probability Density Function}
    \acro{XR}{eXtended Reality}
    \acro{FSM}{Finite State Machine}
    \acro{ITQ}{Immersive Tendencies Questionnaire}
    \acro{BFI}{Big Five Inventory of personality traits}
    \acro{PCA}{Principal Component Analysis}
    \acro{ECDF}{Empirical \ac{CDF}}
    \acro{NSF}{United States National Science Foundation}
\end{acronym}
\newcommand*\dif{\,\mathrm{d}}

\usepackage[
  style=numeric-comp,
  sorting=none,
  sortcites,
  hyperref,
  mincitenames=1,
  maxcitenames=2,
  maxbibnames=2,
  minbibnames=1,
  citestyle=numeric-comp, 
  backend=bibtex
]{biblatex}
\bibliography{bibliography.bib}
\AtBeginBibliography{\small}
\AtEveryBibitem{\clearfield{day}}
\AtEveryBibitem{\clearfield{isbn}}
\AtEveryBibitem{\clearfield{series}}
\AtEveryBibitem{\clearlist{location}}
\AtEveryBibitem{\clearfield{doi}}

\usepackage{orcidlink}
\usepackage[all]{hypcap}
\usepackage{makecell}
\usepackage[capitalize,nameinlink,noabbrev]{cleveref}

\crefname{appendixwithoutnumber}{appendix}{appendices}
\crefformat{appendixwithoutnumber}{#2the \appendixname#3}
\Crefformat{appendixwithoutnumber}{#2The \appendixname#3}

\usepackage[inline]{enumitem}
\setlist[description]{font=\normalfont\underline}

\hypersetup{
  hidelinks,
  colorlinks=true,
  allcolors=black,
  pdfstartview=Fit,
  breaklinks=true
}

\newcommand{\edgedroid}{EdgeDroid~\num{2.0}}

\begin{document}
\title{Realistic modeling of human timings for \acl*{WCA}}

\author{%
  \IEEEauthorblockN{%
    \begin{tabularx}{\textwidth}{@{}XXX@{}}
      \makecell{%
        Manuel {Olguín Muñoz}\IEEEauthorrefmark{1}~\orcidlink{0000-0002-3383-2335}%
      } &
      \makecell{%
        Vishnu N. Moothedath\IEEEauthorrefmark{2}~\orcidlink{0000-0002-2739-5060}%
      } &
      \makecell{%
        {Jaya Prakash} Champati\IEEEauthorrefmark{3}~\orcidlink{0000-0002-5127-8497}%
      }\\
      \makecell{%
        Roberta Klatzky\IEEEauthorrefmark{4}~\orcidlink{0000-0001-9701-9186}%
      } &
      \makecell{%
        Mahadev Satyanarayanan\IEEEauthorrefmark{5}~\orcidlink{0000-0002-2187-2049}%
      } &
      \makecell{%
        James Gross\IEEEauthorrefmark{6}~\orcidlink{0000-0001-6682-6559}%
      }%
    \end{tabularx}
  }\\\vspace{0.5em}
  \begin{tabularx}{\textwidth}{@{}XXX@{}}
    \makecell[t]{%
      \IEEEauthorblockA{%
        \IEEEauthorrefmark{1}\IEEEauthorrefmark{2}\IEEEauthorrefmark{6}EECS School\\
        KTH Royal Institute of Technology\\Sweden\\
        \{\IEEEauthorrefmark{1}\href{mailto:molguin@kth.se}{molguin}, 
        \IEEEauthorrefmark{2}\href{mailto:vnmo@kth.se}{vnmo},
        \IEEEauthorrefmark{6}\href{mailto:jamesgr@kth.se}{jamesgr}\}\\@kth.se
      }
    } & 
    \makecell[t]{%
      \IEEEauthorblockA{%
        \IEEEauthorrefmark{3}Edge Networks Group\\
        IMDEA Networks Institute\\Spain\\
        \href{mailto:jaya.champati@imdea.org}{jaya.champati@imdea.org}
      }
    } & 
    \makecell[t]{%
      \IEEEauthorblockA{%
        \IEEEauthorrefmark{4}Department of Psychology\\ 
        \IEEEauthorrefmark{5}School of Computer Science\\
        Carnegie Mellon University\\USA\\
        \IEEEauthorrefmark{4}\href{mailto:klatzky@cmu.edu}{klatzky@cmu.edu}\\
        \IEEEauthorrefmark{5}\href{mailto:satya@cs.cmu.edu}{satya@cs.cmu.edu}
      }
    }
  \end{tabularx}\\\vspace{0.5em}
}
\maketitle

\begin{abstract}
    \ac{WCA} applications present a challenge to benchmark and characterize due to their human-in-the-loop nature.
    Employing user testing to optimize system parameters is generally not feasible, given the scope of the problem and the number of observations needed to detect small but important effects in controlled experiments.
    Considering the intended mass-scale deployment of \ac{WCA} applications in the future, there exists a need for tools enabling human-independent benchmarking.

    We present in this paper the first model for the complete end-to-end emulation of humans in \ac{WCA}.
    We build this model through statistical analysis of data collected from previous work in this field, and demonstrate its utility by studying application task durations.
    Compared to first-order approximations, our model shows a \textasciitilde\SI{36}{\percent} larger gap between step execution times at high system impairment versus low.
    We further introduce a novel framework for stochastic optimization of resource consumption-responsiveness tradeoffs in \ac{WCA}, and show that by combining this framework with our realistic model of human behavior, significant reductions of up to \SI{50}{\percent} in number processed frame samples and \SI{20}{\percent} in energy consumption can be achieved with respect to the state-of-the-art.
\end{abstract}
\acresetall%

\section{Introduction}\label{sec:intro}

\acf{WCA} applications are a novel category of wearable, edge-native applications aiming to amplify human cognition in both daily activities and professional settings.
These systems aim to seamlessly integrate into the day-to-day of users, leveraging compute-intensive algorithms to analyze user and environment information and provide real-time, context-aware information and feedback.
\ac{WCA} applications originally emerged as assistive use-cases for individuals suffering from cognitive decline due to aging or traumatic brain injuries~\cite{satyanarayanan2009case,Ha2014towards,Satya2019augmenting}, and have since expanded to a greater range of use cases.
In particular, following the success of non-wearable \ac{XR} and cognitive assistance in industrial settings~\cite{Funk2015Cognitive,Wang2022Comprehensive}, there is increasing interest in the research community in the application of \ac{WCA} for step-by-step assistance for complex assembly tasks~\cite{Chen2017Empirical,belletier2021wearable}.

A defining characteristic of these applications is their lack of reliance on intentional user inputs to trigger responses.
They are intended to operate as autonomous guides, much akin to how \ac{GPS} systems guide drivers, tracking their progress and providing feedback and instructions at appropriate times.
This context-sensitivity and proactivity in providing user feedback translate into a reliance on high-dimensional, complex, unstructured inputs, such as real-time video, which require intensive compute capabilities to process.
This also translates into latency-sensitivity, as with any \ac{AR} application.
Delays and jitter can be jarring to the user, causing discomfort and leading them to make mistakes and potentially even abandoning the application altogether.
On the other hand, as their name suggests, these systems are by design \emph{wearable}, which implies the use of lightweight, battery-powered, low energy consumption devices.

The combination of these opposing characteristics has led to \ac{WCA} applications being identified in the literature as prime candidates for offloading to the edge~\cite{Ha2014towards,Chen2017Empirical,Chen2018application}.
However, many unknowns still remain before consumer-scale adoption of these applications can become a reality.
One key gap in knowledge pertains to the current lack of tools and methodologies for scalable and repeatable study of \ac{WCA} application performance and resource utilization in realistic deployments.
Due to their human-in-the-loop nature, these applications present a challenge to benchmark and characterize, in particular in real-scale deployments where dozens or even hundreds of users might concurrently use the system.
Accordingly, recruiting a  cohort of subjects for realistic benchmarking and study of \ac{WCA} systems can be prohibitively cumbersome and expensive for many research groups and system designers
There exists therefore a real need for scalable tools for \ac{WCA} benchmarking which do not rely on direct testing of the human-in-the-loop.

\medskip

In that context, the contributions of this paper are:
\begin{enumerate}
    \item\label{item:contrib:model} We introduce the first, to our knowledge, stochastic model for human timings in \ac{WCA} applications.
    Using the data collected for~\cite{olguinmunoz:impact2021} as a base, we build a stochastic model which takes as input past measurements of system responsiveness and produces realistic step execution times.
    We also introduce a novel way to generate dynamic traces of frames for \ac{WCA} applications which can be combined with the timing model for a full end-to-end emulation of a human.
    We name this new model \emph{\edgedroid}; a direct, more realistic evolution of our initial EdgeDroid approach~\cite{olguin2018scaling,olguin2019edgedroid}.
    \item\label{item:contrib:footprint} Using this model, we study the implications of realistic human behavior for the \emph{application lifetime footprint} of \ac{WCA}, understanding this term as the duration of a specific complete execution of the application task.
    In accordance with previous work~\cite{olguinmunoz:impact2021}, we find dependencies between system responsiveness and human step execution times that lead to substantially different application lifetimes when compared to a first-order baseline which does not take into account human behavior.
    \item\label{item:contrib:optimization} Finally, we study the potential for optimization in \ac{WCA} when considering human behavior using our model.
    We develop a generic model for the stochastic optimization of resource consumption versus responsiveness trade-offs in these applications, which we apply to two potential avenues for \ac{WCA} optimization; number of processed samples and energy consumption per step.
    First, we study the potential for reducing the number of samples captured per step.
    This is a valuable endeavor, as reducing the number of samples captured and subsequently processed directly translates in lower bandwidth demand on the wireless network and processor time demand on the cloudlet.
    Next, we explore the potential for direct optimization of energy consumption.
    The economic feasibility of WCA, and hence its likelihood of commercial adoption depends on it not being a resource hog, and this work furthers that effort.
    Leveraging our more involved user model, combined with the introduced optimization approach, we achieve a \textasciitilde\SI{60}{\percent} reduction in the number of samples processed per step.
    We also achieve an improvement of \SI{20}{\percent} in energy consumption, all while maintaining comparable levels of system responsiveness.
\end{enumerate}

\medskip

This paper is structured as follows.
\cref{sec:relwork} discusses related work in the field of \acl{WCA} modeling.
In \cref{sec:background} we define and discuss key concepts in \ac{WCA}, as well as summarize the key conclusions from our previous work relating to the relationship between system responsiveness and human behavior in these applications~\cite{olguinmunoz:impact2021}.
In \cref{sec:model} we detail our model for the generation of realistic human timing in \ac{WCA}.
We present its design, verify its expected behavior with respect to our previous results, and introduce the dynamic trace generation for full end-to-end emulation of human behavior.
Next, in \cref{sec:implications:footprint}, we discuss the potential implications of such a model on application lifetimes by studying a small series of representative scenarios.
This is followed by a more in-depth investigation in \cref{sec:implications:optimization} on the potential consequences of a human timing model on the general optimization of \ac{WCA} systems.
In the same section, we introduce our generic optimization framework for resource consumption versus responsiveness trade-offs in \ac{WCA}.
Finally, in \cref{sec:conclusion} we summarize and conclude this paper, as well as briefly discuss potential avenues for future work.
\section{Related work}\label{sec:relwork}

There exist a number of approaches to the characterization and benchmarking of mobile \ac{XR}, and by extension \ac{WCA}, deployments.
\emph{OpenRTiST}~\cite{george2020openrtist} is a tool which uses a compute-heavy, yet latency-sensitive workload --- image style transfer~\cite{jing2019neural} --- to load and benchmark edge computing deployments.
It acts as an end-to-end realistic \ac{AR} workload, allowing thus researchers to measure the real end-to-end latencies of such systems.
\textcite{Chen2017Empirical} use a number of prototype \ac{WCA} implementation to study and characterize real-world latency bounds in these systems.
In~\cite{lecci2021open}, the authors collect and analyze large amounts of \ac{VR} network traffic, which they then use to construct a synthetic model for the generation of traces of such traffic.
\textcite{chetoui2022arbench} propose \emph{ARBench}, a toolkit for the benchmarking of mobile hardware in the context of \ac{AR}.
The toolkit incorporates a series of workloads which stress different components of the mobile device and calculates a score for each workload.
\emph{Yahoo Cloud Serving Benchmark}~\cite{cooper2010benchmarking} and \emph{DCBench}~\cite{jia2013characterizing} are workload sets intended for benchmarking cloud services which have further been used in the context of benchmarking edge computing infrastructure for \ac{AR} suitability.
\emph{Edgebench}~\cite{das2018edgebench} and \emph{Defog}~\cite{mcchesney2019defog} benchmark workload performance on the edge versus on the cloud.

We contribute to this body of work by providing the first tool for the benchmarking of \ac{WCA} that dynamically considers the effects of human behavior.
In contrast to the above described works, which generate loads with relatively static profiles, our model is able to react to changes in system responsiveness like a human would.
We achieve this by building upon our own previous contributions to this field.
In~\cite{olguin2018scaling,olguin2019edgedroid} we introduced a coarse approximation to human behavior modeling in \ac{WCA} we called \emph{EdgeDroid}.
We used a trace-driven approach, where a pre-recorded and pre-processed ``ideal'' trace of steps for a specific \ac{WCA} step-based task is replayed to a Gabriel~\cite{Chen2018application} backend.
In order to adapt to potential mismatches between system responsiveness at trace-capture time and trace-replay time, we used a simple \ac{FSM} to adapt the trace at the latter by replaying or skipping certain segments.
In~\cite{olguinmunoz:impact2021} we studied the effects of reduced system responsiveness on human behavior in \ac{WCA} applications through human-subject studies.
We found that humans generally pace themselves according to the perceived system responsiveness.
When interacting with a highly responsive system, humans tended to speed up with each step; conversely, humans tended to slow down in highly unresponsive states.

The characterization and optimization of mobile \ac{XR} has also long been a topic of research.
\textcite{srinivasan2009performance} analyze a \ac{MAR} workload and identify and characterize the computational bottlenecks in the application.
They use this characterization to develop a series of code optimizations which allow them to achieve a threefold increase in performance.
\textcite{Wang2019Towards} develop a taxonomy for \ac{WCA} as well as strategies for workload reduction in these applications, including a novel adaptive sampling scheme with the goal of reducing the number of processed samples while maintaining application responsiveness.
\textcite{huang2021proactive} present a framework for the proactive allocation of edge cloud resources while taking into account the inherent mobility of wearable and mobile \ac{AR}.
\textcite{al2017energy} propose a resource allocation strategy which leverages collaboration between \ac{AR} applications to minimize energy consumption.
As above, we have also previously directly contributed to this field.
In~\cite{Moothedath2021EnergyOptimal,Moothedath2022EnergyEfficient} we find the optimum periodic sampling interval which minimizes the energy tracking human progress of a specific subtask in \ac{WCA}.
This is extended into an aperiodic strategy in~\cite{Moothedath2022Aperiodic}.
We further develop this approach in this work by deriving a more practical approximate solution to the aperiodic strategy, which we then use to implement a novel framework for the generic optimization of responsiveness-resource consumption trade-offs.
\section{Background of \acs*{WCA}}\label{sec:background}

\begin{figure}
    \centering
    \begin{subfigure}{\columnwidth}
        \centering
        \includegraphics[width=\columnwidth]{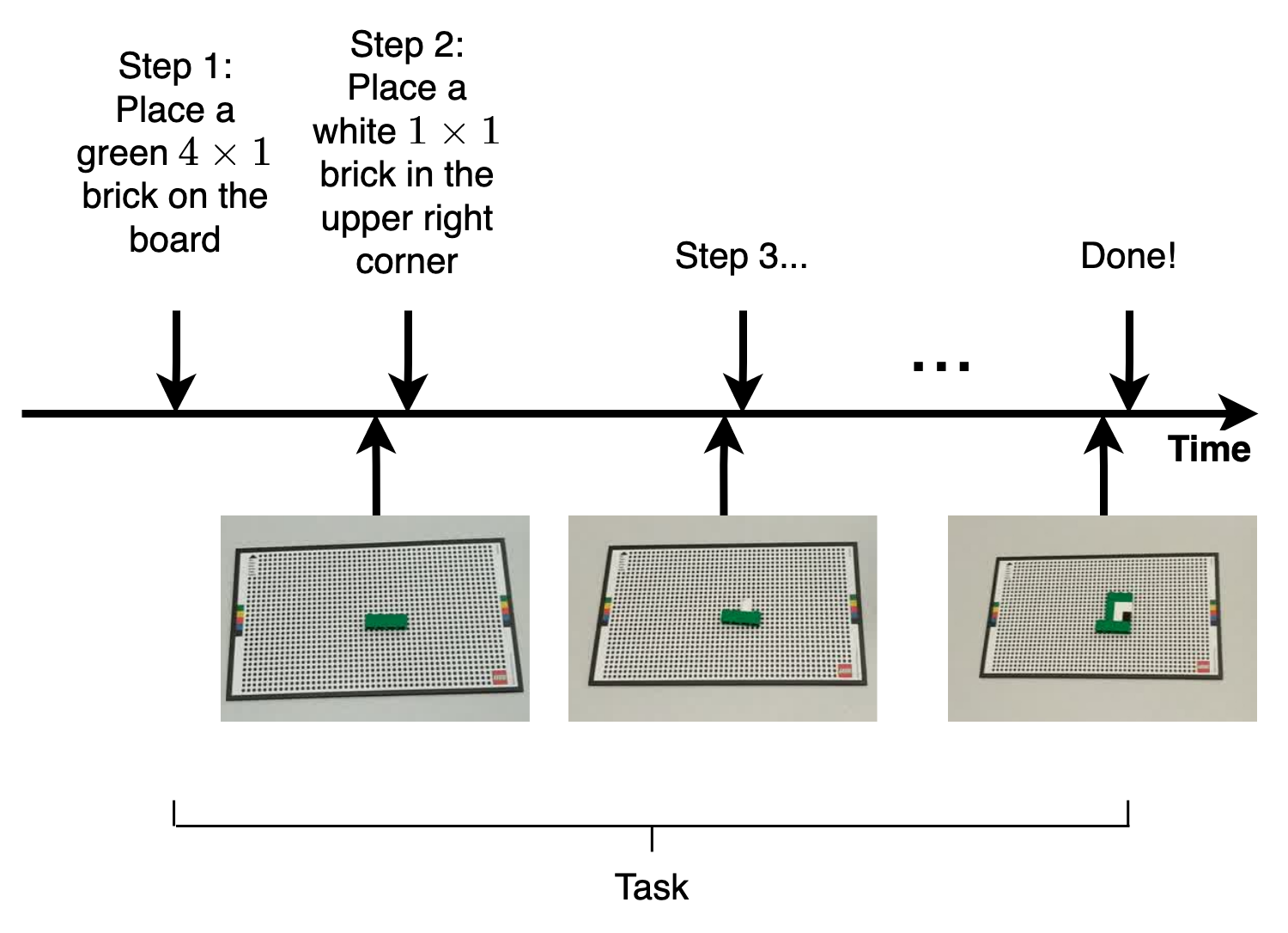}
        \caption{%
            Overview of a task in a \ac{WCA}, composed of a series of steps.
            Each steps starts with an instruction being provided to the user and ends with the instruction for the next step.
            The \ac{WCA} continuously samples the task state, automatically triggering transitions between steps as correct (or incorrect) states are recognized.
        }\label{fig:task}
    \end{subfigure}\\
    \begin{subfigure}{\columnwidth}
        \centering
        \includegraphics[width=\columnwidth]{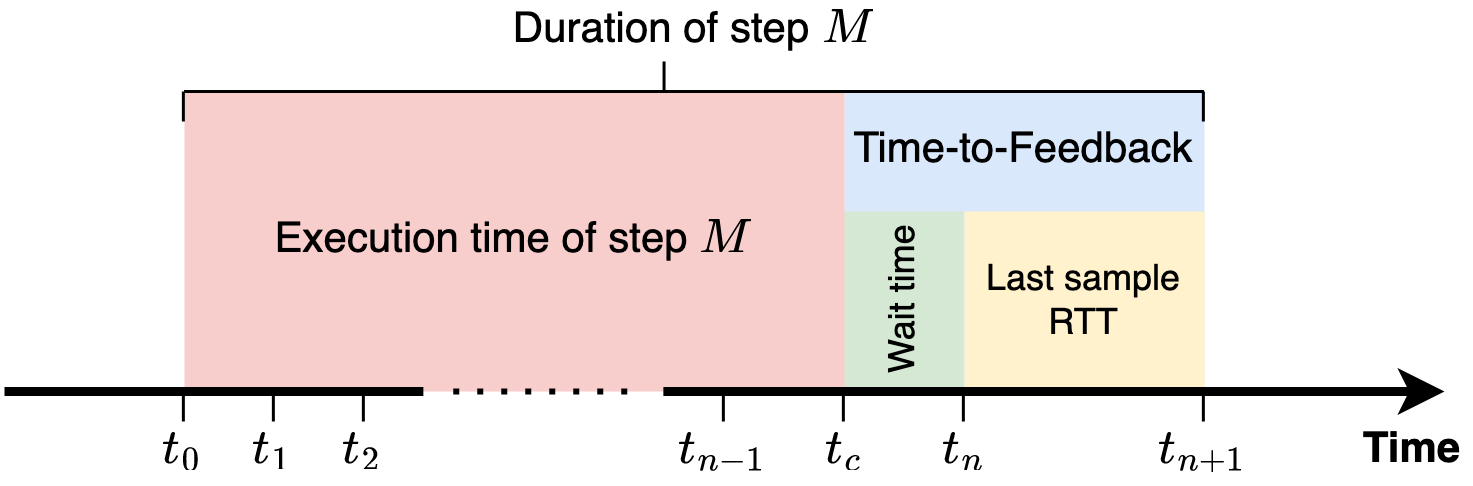}
        \caption{%
            Breakdown of a step into its timing components.
            The instruction for step \( M \) and \( M + 1 \) are provided to the user at \( t_0 \) and \( t_{n+1} \), respectively.
            \( t_k | k \in \{1, \ldots, n \} \) correspond to the \ac{WCA} sampling instants for step \( M \), and \( t_c \) marks the instant at which the user finishes performing the instruction.
        }\label{fig:step}
    \end{subfigure}
    \caption{Key concepts in \acl{WCA}}
\end{figure}

\acf{WCA} applications represent a category of novel, context-sensitive and highly-interactive \ac{AR} applications.
In this work we focus on a particular category --- ``step-based'' \ac{WCA} --- that have as their goal the guiding of a user through a sequential task.
Examples of such applications are the LEGO and IKEA assistants~\cite{Chen2015LEGO,Chen2018application}, in which users are guided step-by-step through the process of assembling a LEGO model and an IKEA lamp, respectively.

Step-based \acp{WCA} operate analogously to how \ac{GPS} navigation assistants guide users, by seamlessly and continuously monitoring the progress of the user and autonomously providing relevant instructions and feedback.
The application follows the progress of the task in ``realtime'' by repeatedly sampling the state of the physical system, most commonly through video frames.
Whenever the assistant detects that the user has correctly or incorrectly performed an instruction, it provides a new instruction to either advance the task or correct the detected mistake.
The application otherwise remains silent and out-of-the-way of the user; that is, samples which do not generate a new instruction (e.g.~because they captured an intermediate or unfinished state, or simply noise) are silently discarded.
Herein lies one of the key characteristics of these applications: the user only consciously interacts with the application whenever they finish an instruction, and thus these are the \emph{only} points in time at which they can notice changes in system responsiveness.

In order to discuss these applications with precision we provide some definitions relating to their operation.
First of all, a \emph{step} is formally understood as a specific action to be performed by the user, described by a single instruction, and a \emph{task} consists of a series of steps to be executed in sequence (see \cref{fig:task}).
A step begins when the corresponding instruction is provided to the user, and ends when the instruction for the next step is provided; we call the time interval between these two events the \emph{step duration}.

\acp{WCA} employ sampling, most commonly of video feeds, to keep track of the state of the real world.
Take \( \{ t_0, t_1, \ldots, t_{n + 1} \} \) a series of discrete and sequential sampling instants at which the \ac{WCA} captures the state of the physical system, as illustrated in \cref{fig:step}.
\( t_0 \) corresponds to the instant at which the instruction for step \( M \) is provided and the first sample is taken, and \( t_n \) to the instant at which the final sample (i.e.\ which captures the final state of step \( M \)) is taken. 
\( t_{n + 1} \) is then the instant at which the result for sample \( t_n \) is returned and the instruction for step \( M + 1 \) is provided.
We define \( t_c \) as the point in time at which the user finishes performing the instruction for step \( M \), and the intervals \( t_c - t_0 \), \( t_n - t_c \), and \( t_{n + 1} - t_n \) as the \emph{execution time}, \emph{wait time}, and \emph{last sample \ac{RTT}}, respectively, of a step \( M \).
The sum of the latter two values (i.e.\ the interval \( t_{n + 1} - t_c \)) we call \emph{\ac{TTF}}, a metric which we will repeatedly refer to in this paper as it directly describes the responsiveness of a \ac{WCA}.



\subsection{The effects of changes in responsiveness on human behavior}\label{ssec:plos}

In~\cite{olguinmunoz:impact2021} we studied the effects of system responsiveness on human behavior in step-based \ac{WCA} in a controlled experiment.
We employed a modified and instrumented version of the above-mentioned, step-based LEGO \ac{WCA}~\cite{Chen2015LEGO}.
\num{40} subjects interacted with the assistant, performing a \num{169}-step task while we altered the responsiveness in realtime and captured key application and task performance metrics.
Additionally, we also employed questionnaires to evaluate personality traits of the participants, and correlated these results with the task performance metrics.

We used \emph{delay} and execution time as the variables for system responsiveness and human behavior, respectively.
\emph{Delay} corresponded to a temporarily fixed time duration for the processing interval of each frame.
That is, if during a series of steps delay was set to \( D \) seconds, the feedback for each frame was provided to the user \( D \) seconds after frame capture.
The correlation between this variable and execution time was then studied.
To compensate for the distribution of \( t_c \)\footnote{\( t_c \), and conversely the wait time \( \mathcal{W} \) of a step, can be assumed to be uniformly distributed in the interval \( [t_{n - 1}, t_n] \) without loss of generality.}~\cite{olguinmunoz:impact2021}, in the present work we adjust the nominal delay \( D \) by a factor \num{1.5}.
We will refer to this adjusted delay as \emph{\acf{TTF}}.




Our findings can be summarized as follows:
\begin{enumerate}
    \item System slow-down induces \emph{additional} behavioral slow-down which scales with the decrease in system responsiveness.
    Compared to the unimpaired case, participants were on average \SI{12}{\percent} slower when subject to a mean \ac{TTF} of \SI{2.475}{\second}, and \num{26} percentage points slower at a mean \ac{TTF} of \SI{4.5}{\second}.

    \item\label{item:speedup} At lower \acp{TTF}, humans get faster at performing steps as the task progresses, 
    For sequences of \num{12} steps in an unimpaired application state, humans executed the final four steps on average \SI{36}{\percent} faster than they did the first four.
    However, this effect is dampened by reduced system responsiveness, and actually inverts at the highest levels of system impairment; humans actually become progressively slower the longer they spend in a degraded system state.

    \item\label{item:remain} The effects of system slow-down on human behavior remain for a while even after system responsiveness improves.
    These effects are noticeable for at least \num{4} steps after the return to a high-responsiveness state. 
    
    \item The above effects are modulated by measures of individual levels of personality characteristics and focus.
\end{enumerate}

\subsubsection{Moderation of effects by individual characteristics}\label{ssec:moderationeffects}

We recorded variables related to well-known individual differences, encompassing both the \acf{BFI}~\cite{oliver:bfi1999}, and the \acf{ITQ}~\cite{witmer1998measuring}.
Out of the individual-difference variables, the most salient effect on performance corresponded to \emph{neuroticism}, a \ac{BFI} trait linked to low tolerance for stress and high emotional reactivity, and which has previously been linked to higher \emph{delay discounting} rates~\cite{hirsh2008delay}.
Delay discounting is the tendency to devalue rewards for which one must wait; high rates, indicative of waiting intolerance, have been associated with negative social and academic outcomes.


In this work, we will use a normalized scale to describe neuroticism, derived from the minimum and maximum obtainable values for this variable in the \ac{BFI}~\cite{oliver:bfi1999}.
\emph{Low} and \emph{high} neuroticism will refer to the \( [0.0, 0.5) \) and \( [0.5, 1.0] \) ranges respectively.

Linear regression showed a significant correlation between individual neuroticism scores and the execution time late in a series of high-delay steps, \( \rho = 0.418 \), \num{2}-tailed \( p < 0.05 \)
Neuroticism was further identified as a modulating factor for the pacing effects through a \ac{PCA}.
Out of the three identified components, which cumulatively accounted for \SI{73.13}{\percent} of the variance in the results, neuroticism was included in the first two.
The effect of neuroticism was observed across all \acp{TTF} and impairment durations in the tasks.

\begin{figure}
    \centering
    \includegraphics[width=\columnwidth]{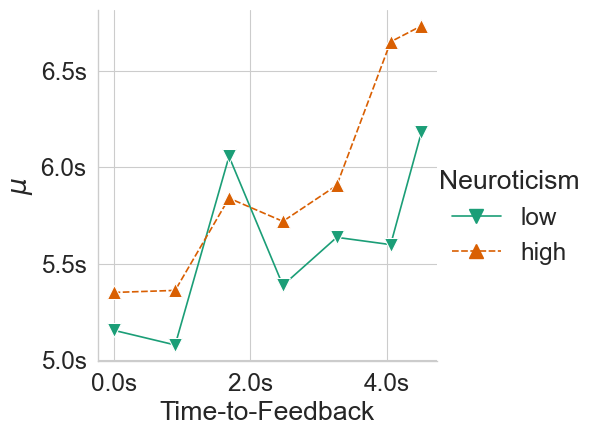}
    \caption{%
        \( \mu \) parameter of \acs*{exGaussian} distributions fitted to execution times of the first four steps of segments of steps subject to the same \ac{TTF} in~\cite{olguinmunoz:impact2021}.
        Distributions were fitted using \ac{MLE}.
    }\label{fig:muexgaussian}
\end{figure}

\begin{figure*}
    \centering
    \includegraphics[width=\textwidth]{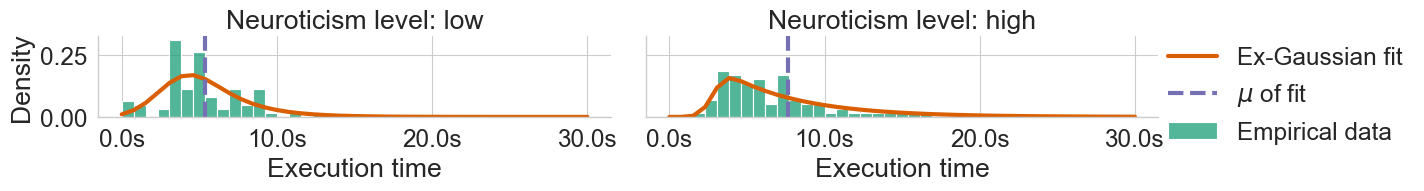}
    \caption{%
        Example \ac{exGaussian} fits on execution times from steps \numrange{4}{8} in a segment of steps at the maximum experimental \ac{TTF}.
        The effects of neuroticism are clearly visible in the tail and the mean of the distributions.
    }\label{fig:fitsneuro}
\end{figure*}

Furthermore, we found that execution times, when grouped by experimental variables such as neuroticism, \ac{TTF}, and continuous segments of steps subject to the same \ac{TTF}, were well-fit by an \ac{exGaussian} distribution, as verified using Kolmogorov-Smirnov goodness-of-fit tests~\cite{massey1951kolmogorov}.
When grouping by level of neuroticism, \ac{TTF}, and \emph{slice}\footnote{%
In~\cite{olguinmunoz:impact2021} the \emph{slice} to which a step belongs to refers to whether the step occurred in the first, second, or third four-step segment of a sequence of steps subject to the same \ac{TTF}.
}, the best fit statistic was \ensuremath{0.028} (\ensuremath{p = 0.999}).
This distribution has an ample body of research supporting its suitability for the modeling of the timing of human actions and reaction times~\cite{Rohrer1994analysis,Palmer2011shapes,Marmolejo2022generalised}.
We found that the effects of neuroticism on execution times were clearly identifiable in the fitted distributions, in particular in their means and tails.
\cref{fig:muexgaussian} shows an example of this modulating effect, illustrating the behavior of the mean (\( \mu \) parameter) of \ac{exGaussian} distributions fitted to the execution times of the first four steps of segments of steps subject to the same \ac{TTF}.
Finally, \cref{fig:fitsneuro} shows an example of the effects of neuroticism on the fitted \ac{exGaussian} distributions for a specific group of execution times.
Higher neuroticism directly translates into a higher mean and longer tail.
\section{A model of human behavior for \ac{WCA}}\label{sec:model}

In the following, we employ the above insights together with the data collected for~\cite{olguinmunoz:impact2021} to design and build a probabilistic model of human behavior for \ac{WCA}.
We detail the construction of a model which uses the data collected to generate, at runtime, realistic execution times. 
In order to accurately emulate the behavior of a human, such a model needs to implement two main behaviors.
First, it needs to generate realistic execution times for each step in the task, considering the current and historical impairment of the \ac{WCA} system, as well as salient individual-difference measures.
We detail this in \cref{ssec:model:exectimes}.
Neuroticism is incorporated in this aspect of the model, as it is the most salient individual difference in the data; however, other factors such as immersive tendency could be treated similarly.
Second, the model must produce sequences of input samples for each step mimicking what a real human would generate; this is explained in \cref{ssec:model:frames}. 

\subsection{Generating realistic execution times}\label{ssec:model:exectimes}

The processing of the data from~\cite{olguinmunoz:impact2021} for the generation of execution times can be summarized as a grouping according to discretized levels of neuroticism and a weighted rolling average of \ac{TTF}.
The resulting collections of execution times represent the distributions of these values for users with specific levels of neuroticism, interacting with systems at specific states of impairment and recent histories of impairment.
These distributions can then be sampled to produce new, realistic execution times.
In the following, we detail the step-by-step processing of this data and construction of the probabilistic model.

We employ a cleaned and re-parameterized copy of the timing data.
The \num{6760} data points are arranged in a table together with identifiers for the subjects, their normalized neuroticism score, and a sequence number for each step.

\begin{table*}[]
    \centering
    \caption{%
        Default bins used for model parameter levels.
        Values have been rounded to two decimal places.
    }
    \label{tab:defaultbins}
    \begin{tabular}{@{}lccccccc@{}}
        \toprule
        \textbf{Parameter} & \textbf{Low} & & & \textbf{Medium} & & & \textbf{High}         \\ \midrule
        Neuroticism        & \( [0, 0.5) \) & & & & & & \( [0.5, 1.0] \)      \\
        Weighted TTF & \([0.0], 0.82]\) & \((0.82, 1.53]\) & \((1.53, 2.08]\) & \((2.08, 2.67]\) & \((2.67, 3.45]\) & \((3.45, 4.13]\) & \((4.13, \infty]\)
    \end{tabular}
\end{table*}

We begin by calculating rolling weighted averages of the \acp{TTF} of each of the \num{40} individual repetitions of the data.
We use exponentially decaying weights for the most recent \num{12} steps, defined in \cref{eq:weights}, such that the most recent \ac{TTF} accounts for roughly \SI{50}{\percent} of the rolling average, the second-most recent for \textasciitilde\SI{25}{\percent}, the third-most \textasciitilde\SI{12}{\percent}, and so on.
We additionally pad the data for each run with \num{12} copies of the first \ac{TTF} in order to ensure sensible values for the first twelve steps; this padding is removed after the weighted values have been calculated.
This weighing ensures that drastic changes in \ac{TTF} are significantly remembered by the model for at least \num{4} steps --- in line with our previous findings on human behavior --- and are subsequently quickly forgotten.
\begin{equation}\label{eq:weights}
    w_{n - i} = 
    \left\{ \begin{array}{ll}
        \frac{e^{-0.7 i}}{\sum\limits^{12}_{j=1} e^{-0.7 j}} & 1 \leq i \leq 12 \\
        & \\
        0 & i > 12
    \end{array} \right.
\end{equation}

The resulting weighted \acp{TTF} are then binned into continuous ranges by splitting the data on the \num{7}-quantiles.
The exact resulting bins of this operation on the data are presented in \cref{tab:defaultbins}.

Next, the data is further tagged according to its associated level of normalized neuroticism.
For this work, we use two levels, \emph{low} and \emph{high}, the exact values for which can also be seen in \cref{tab:defaultbins}.

After this preprocessing has been finished, the data is ready to be used for the generation of execution times by applying the above steps in real-time to measured \acp{TTF}:

\begin{enumerate}
    \item At the beginning of each step, the model is fed the measured \ac{TTF} for the previous step.
    \item The model calculates a weighted average over the latest \num{12} steps using the weights defined in \cref{eq:weights}.
    \item The resulting weighted \ac{TTF} is binned into the corresponding range (\cref{tab:defaultbins}).
    \item The model then filters the pre-processed data to find execution time samples associated with this same discretized weighted \ac{TTF}.
    \item The selected samples are further filtered to match the desired level of neuroticism.
    \item The remaining samples are then used to output a realistic execution time, either by
    \begin{itemize}
        \item directly sampling the execution time values;
        \item or, using \ac{MLE} to fit a distribution to the execution time samples and then sampling the distribution instead.
    \end{itemize}
\end{enumerate}

The distribution chosen for the second variant described above corresponds to the \ac{exGaussian} distribution, previously discussed in \cref{ssec:moderationeffects}.

We implement these two variants of the model in Python 3.10, and verify their correct behavior in the following.

\subsubsection{Verifying the behavior of the timing model}\label{ssec:model:verification}

We verify the behavior of the above described model with respect to the four main conclusions of our work in~\cite{olguinmunoz:impact2021}, which were previously discussed in \cref{ssec:plos}.
These are reformulated as objectives below:

\begin{enumerate}
    \item\label{it:ttftoexectime} Higher \acp{TTF} should result in higher execution times.
    \item\label{it:duration} When subject to a series of steps at the same level of impairment, desired behavior depends on the level:
    \begin{enumerate}
        \item At low \acp{TTF}, the model should speed up; i.e.\ execution times should decrease.
        \item At medium \acp{TTF}, execution times should remain more or less the same.
        \item At high \acp{TTF}, execution times should increase.
    \end{enumerate}
    \item The effects on execution times due to past changes in system responsiveness should linger for at least a few steps whenever system responsiveness changes anew.
    \item\label{it:neuro} Finally, neuroticism should act as a modulating factor for the above effects.
\end{enumerate}

\begin{figure}
    \centering
    \includegraphics[width=\columnwidth]{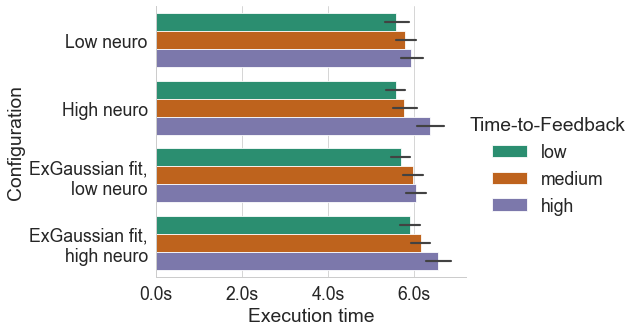}
    \caption{%
        Effects of feeding three different \acp{TTF} (\emph{low}, \SI{0}{\second}, \emph{medium}, \SI{2.5}{\second}, or \emph{high}, \SI{5}{\second}) into the model on the generated execution times.
        Higher \acp{TTF} directly lead to higher execution times.
        Error bars indicate the \SI{95}{\percent} \ac{CI}.
    }\label{fig:ttf_to_exectime}
\end{figure}

\cref{fig:ttf_to_exectime} shows the mean execution time outputted by the model when fed three different levels of \ac{TTF} (\emph{low}, \SI{0}{\second}, \emph{medium}, \SI{2.5}{\second}, or \emph{high}, \SI{5}{\second}).
These results were generated by first warming up the model by feeding it \num{25} \acp{TTF} selected at random from the data before feeding it the desired input \ac{TTF}, and recording the generated execution time.
This procedure is repeated \num{600} times for each configuration and target \ac{TTF}.
The resulting mean execution times match precisely the desired behavior mentioned in \cref{it:ttftoexectime}, with the difference in mean execution times at low versus high \acp{TTF} reaching \SI{14}{\percent} (\textasciitilde\SI{5.6}{\second} to \textasciitilde\SI{6.4}{\second}) in the worst case (high neuroticism configuration).
Additionally, we can observe the effects of neuroticism on generated execution times, as specified in \cref{it:neuro}.
At low neuroticism, the average difference between execution times at low versus high \acp{TTF} was of roughly \SI{6.2}{\percent}, compared to \SI{12.5}{\percent} at high neuroticism.

\begin{figure}
    \centering
    \includegraphics[width=\columnwidth]{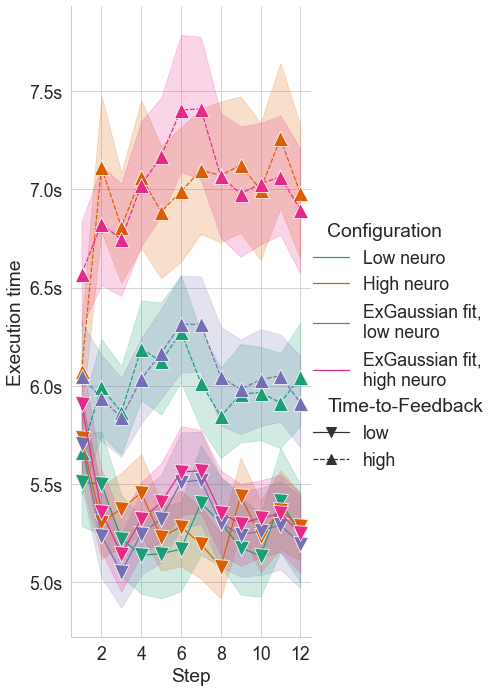}
    \caption{%
    Effects of prolonged exposure to constant levels of system impairment on the model.
    At low (\SI{0}{\second}) \ac{TTF}, the models speed-up over time; conversely at high (\SI{5.0}{\second}) \ac{TTF}, the models either present no change or drastically increase their generated execution times, depending on the level of neuroticism.
    Error bands indicate \SI{95}{\percent} \ac{CI}.
    }\label{fig:exectimeduration}
\end{figure}

Next, \cref{fig:exectimeduration} shows the evolution of generated execution times while the model is subject to a fixed \ac{TTF}, either \emph{low} (\SI{0}{\second}) or \emph{high} (\SI{5}{\second}).
These results were generated by first warming up the model with \num{25} random \acp{TTF}, and then recording the generated execution times over a sequence of \num{12} steps at a fixed \ac{TTF}; this procedure is repeated \num{600} times for each configuration and target \ac{TTF}.
Once again, we see here behavior matching what is expected of the model, in particular with respect to \cref{it:duration}.
At low \acp{TTF}, the model is on average, across all configurations, \SI{8.2}{\percent} faster at step \num{12} when compared to step \num{1}.
At high \acp{TTF}, the behavior changes depending on the level of neuroticism of the model.
Low neuroticism models basically do not change their execution times, whereas high neuroticism configurations are on average \SI{10}{\percent} slower after \num{12} steps.
This is once again in line with our previous findings, as we had previously concluded that humans tend to speed up during a task, but that this speed-up is hindered and eventually reversed as system responsiveness decreases, and that the strength of this effect is correlated with neuroticism.

\begin{figure*}
    \centering
    \includegraphics[width=\textwidth]{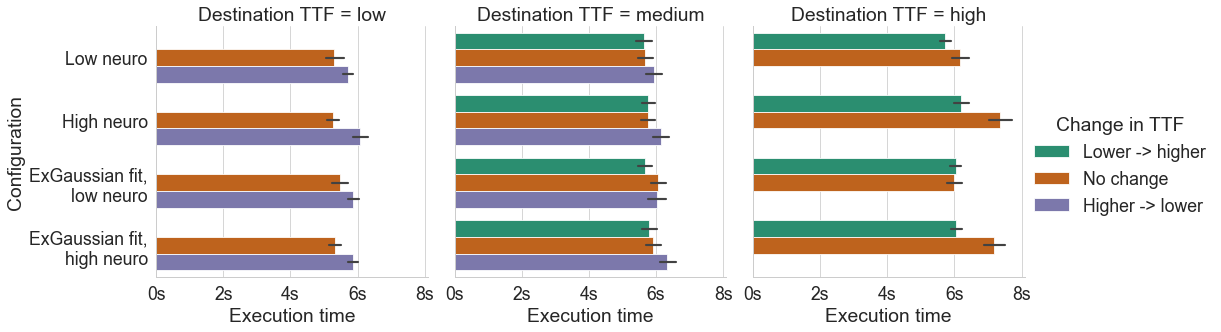}
    \caption{%
        Effects of changes in system impairment on subsequently generated execution times.
        These effects linger on after the change, and thus execution times immediately after a transition are either consistently lower or higher than otherwise at the new \ac{TTF}, depending on the old \ac{TTF}.
        Error bars indicate \SI{95}{\percent} \ac{CI}.
    }\label{fig:transitions}
\end{figure*}

Finally, in \cref{fig:transitions} we showcase the behavior of the model when comparing execution times generated immediately after a change in system responsiveness.
We generate these results by first warming up the model by feeding it a fixed \ac{TTF} (which we will refer to as the \emph{origin} \ac{TTF}) \num{25} times.
Next, another \ac{TTF} value (the \emph{destination} \ac{TTF}) is fed to the model, and we record the output execution time.
Each sample is tagged according to the relation between origin and destination \ac{TTF}, either lower to higher, higher to lower, or equal.
As before, we run \num{600} repetitions of this procedure for each combination of model configuration, origin \ac{TTF}, and destination \ac{TTF}.
Execution times generated immediately after a transition from a higher \ac{TTF} into a lower one are consistently higher than execution times generated without a preceding change in \acp{TTF}.
Conversely, execution times are consistently lower than otherwise immediately after a change from a lower \ac{TTF} into a higher one.
These results are once again in line with our findings in~\cite{olguinmunoz:impact2021}, in which we found lingering effects of transitions between levels of system impairment on human execution times.

\subsection{Generating realistic samples}\label{ssec:model:frames}

Apart from the aforementioned timing and performance data, for~\cite{olguinmunoz:impact2021} we also recorded all collected video frame samples together with matching metadata.
For each video frame submitted to the \ac{WCA} during the tasks, we recorded
\begin{enumerate*}[itemjoin={{; }}, itemjoin*={{; and }}]
    \item the raw video frame captured
    \item sample submission timestamp
    \item \ac{WCA} processing completed timestamp
    \item result or acknowledgement returned timestamp
    \item a tag representing the result of the \ac{WCA} processing
\end{enumerate*}.
The tags assigned corresponded to:
\begin{description}[font={\bfseries\ttfamily}, wide]
    \item[SUCCESS:] frames which triggered a transition to a new step (or the correction of a previous mistake) in the logical task model of the \ac{WCA}, and thus cause the generation of feedback to the user.
    \item[REPEAT:] frames which captured the same board state as the previous successful frame, and thus produced no feedback.
    \item[LOW\_CONFIDENCE] frames for which the image recognition algorithm in the \ac{WCA} did not reach the necessary confidence threshold to interpret it as a valid board state.
        These frames also produce no feedback.
    \item[BLANK] frames in which not enough of the board is visible due to noise, movement, occlusion, etc.
        These frames produce no feedback either.
    \item[TASK\_ERROR] frames which contained an incorrect board state and thus triggered a transition to a procedural corrective step and the generation of feedback to the user.
        However, it must be noted that none of the \num{40} participants made any mistakes during the task, and thus no such frames were encountered in the data.
\end{description}

We correlate this frame data with the step timing data described in \cref{ssec:model:exectimes} to match frames with their corresponding step execution times.
We assign to each frame a normalized instant value \( t_\text{norm} \) corresponding to its capture instant \( \tau \) (expressed in seconds since the start of the step) divided by the total execution time \( t_\text{exec} \) of the step:

\begin{align}
    \left( t_\text{norm} = \frac{\tau}{t_\text{exec}} \right) \in [0, 1]\label{eq:tnorm}
\end{align}

\begin{figure}
    \centering
    \includegraphics[width=\columnwidth]{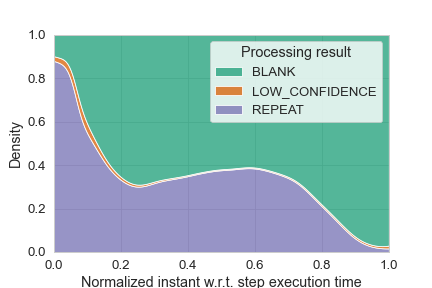}
    \caption{%
        Probability density of frame result tags as a step progresses.
        Note that \texttt{SUCCESS} frames are not included as --- by definition --- the probability for success frames is \num{1} for all normalized instant values greater than or equal to \num{1.0}.
    }\label{fig:frameprobs}
\end{figure}

This allows us to analyze the distribution of frame tag probabilities as a step progresses, independently of execution times.
This is illustrated in \cref{fig:frameprobs}.
Intuitively, \texttt{REPEAT} frames dominate the early instants after a step transition, as the user has not had time to start performing the new instruction and thus the \ac{WCA} keeps capturing frames representing the previous state of the board.
As the user starts moving and performing actions, \texttt{BLANK} frames start to dominate, as this activity prevents the \ac{WCA} from capturing ``clean'' frames.
Finally, it must be noted that \texttt{SUCCESS} frames are not included in this probability density plot, as, by definition:
\begin{equation}
    P(\text{\texttt{SUCCESS}} | t_\text{norm}) =
    \left\{ \begin{array}{ll}
        0 & t_\text{norm} < 1.0    \\
        1 & t_\text{norm} \geq 1.0
    \end{array} \right.
\end{equation}

That is, any frame captured immediately at or after the execution time has been reached will contain the finished board state and thus correspond to a \texttt{SUCCESS} frame.

Using the above insights, together with the corresponding recorded video frames, we devise a scheme for the procedural generation of a synthetic trace for any step in \ac{WCA} task in the same category as those used in~\cite{olguinmunoz:impact2021}.
We first prepare a discretized representation of the probability density map in \cref{fig:frameprobs}.
We segment the normalized instant value into a number of discrete bins (\num{25} in this work), and calculate the relative fraction of frames for each category in each bin.
For each step, given
\begin{enumerate*}[itemjoin={{; }}, itemjoin*={{; and }}]
    \item a collection of random non-\texttt{SUCCESS}, non-\texttt{REPEAT} frames (at least one frame for each of the \texttt{BLANK} and \texttt{LOW\_CONFIDENCE} categories)
    \item an appropriate \texttt{SUCCESS} video frame containing the correct state for the step
    \item an appropriate \texttt{REPEAT} video frame containing the correct state for the \emph{previous} step
\end{enumerate*},
we can then procedurally generate a trace by randomly selecting appropriate frames according to the distributions presented in \cref{fig:frameprobs}.
In other words, for each sampling instant in a step with a given execution time \( t_\text{exec} \):
\begin{enumerate}
    \item We calculate \( t_\text{norm} \) according to \cref{eq:tnorm}.
    \item If \( t_\text{norm} \ge 1.0 \), we select the \texttt{SUCCESS} frame and stop sampling.
    \item If instead \( t_\text{norm} < 1.0 \), we find the appropriate bin for \( t_\text{norm} \) and then select a frame by performing a weighted random sampling of the frame categories in the normalized instant bin.
\end{enumerate}

\subsection{Obtaining the model}\label{ssec:model:obtaining}

We provide the model implementations in Python~\num{3.10} as well as the base data to the community as \ac{FOSS}.
All of these are published on the \href{https://github.com/KTH-EXPECA/EdgeDroid2}{\texttt{KTH-EXPECA/EdgeDroid2}} repository on GitHub under a permissive Apache version 2 license.
\section{Implications for the study of \ac{WCA} footprints}\label{sec:implications:footprint}


We begin by studying the implications of such a model on the estimation of application footprints as described by their lifetimes.
In the context of \ac{WCA}, we will understand \emph{application lifetime} as the time it takes a user to complete a specified task.
This is an important metric for \ac{WCA} optimization, as it directly relates to system resource utilization and contention, and to energy consumption.

In order to illustrate the consequences of using a less realistic model that does not take into account higher order effects, we introduce here a reference model to which we will compare our more realistic models.
This model represents a first-order approximation to empirical execution time modeling, and consist simply of an \ac{exGaussian} distribution fitted to all execution time samples collected for~\cite{olguinmunoz:impact2021}.
This distribution is then randomly sampled at runtime to obtain execution times for each step, without any adjustment to the current state of the system.

We start by studying application lifetimes in a controlled, ideal setup by using the timing models to generate execution times for sequences of \num{100} steps subject to constant \acp{TTF}.
These runs are completely simulated and no sampling of video frames is performed; for each step, we simply feed the models a predefined \ac{TTF} and record the generated execution time.
We use the combination of \acp{TTF} and execution times to calculate theoretical step duration times and subsequent total application lifetimes.
This is done for \num{25} linearly distributed \acp{TTF} in the \SIrange[]{0}{5}{\second} range; \num{45} independent repetitions for each combination of model configuration and \ac{TTF}.

\begin{figure}
    \centering
    \begin{subfigure}[]{\columnwidth}
        \centering
        \includegraphics[width=\textwidth]{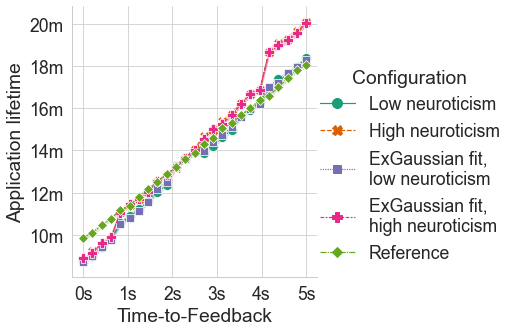}
        \caption{%
            Evolution of mean application lifetimes as \acp{TTF} increase.
            Error bars indicate \SI{95}{\percent} \acp{CI}.
        }
    \end{subfigure}
    \begin{subfigure}[]{\columnwidth}
        \centering
        \includegraphics[width=\textwidth]{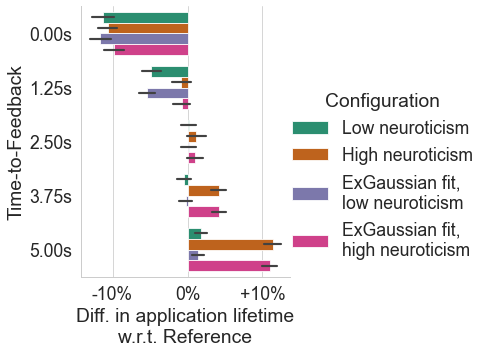}
        \caption{%
            Percentage difference in mean application lifetimes with respect to the reference model at select \acp{TTF}.
            Error bars indicate the \SI{95}{\percent} \acp{CI}, calculated using a two-sided T-test.
        }
    \end{subfigure}
    \caption{\acl{TTF} versus application lifetime.}\label{fig:lifetimes}
\end{figure}

The results of this investigation are presented in \cref{fig:lifetimes}.
Compared to the reference model, our realistic model is, on average, roughly \SI{11}{\percent} faster when subject to low \acp{TTF}.
At higher \acp{TTF}, the behavior of the model depends on its level of normalized neuroticism.
At low neuroticism, the behavior of the realistic model results in total task durations that are basically indistinguishable from the reference model.
However, at high neuroticism, the model once again results in a considerable difference in total task duration with respect to the reference --- this time extending durations by \textasciitilde\SI{11}{\percent} on average.

\begin{figure}
    \centering
    \includegraphics[width=\columnwidth]{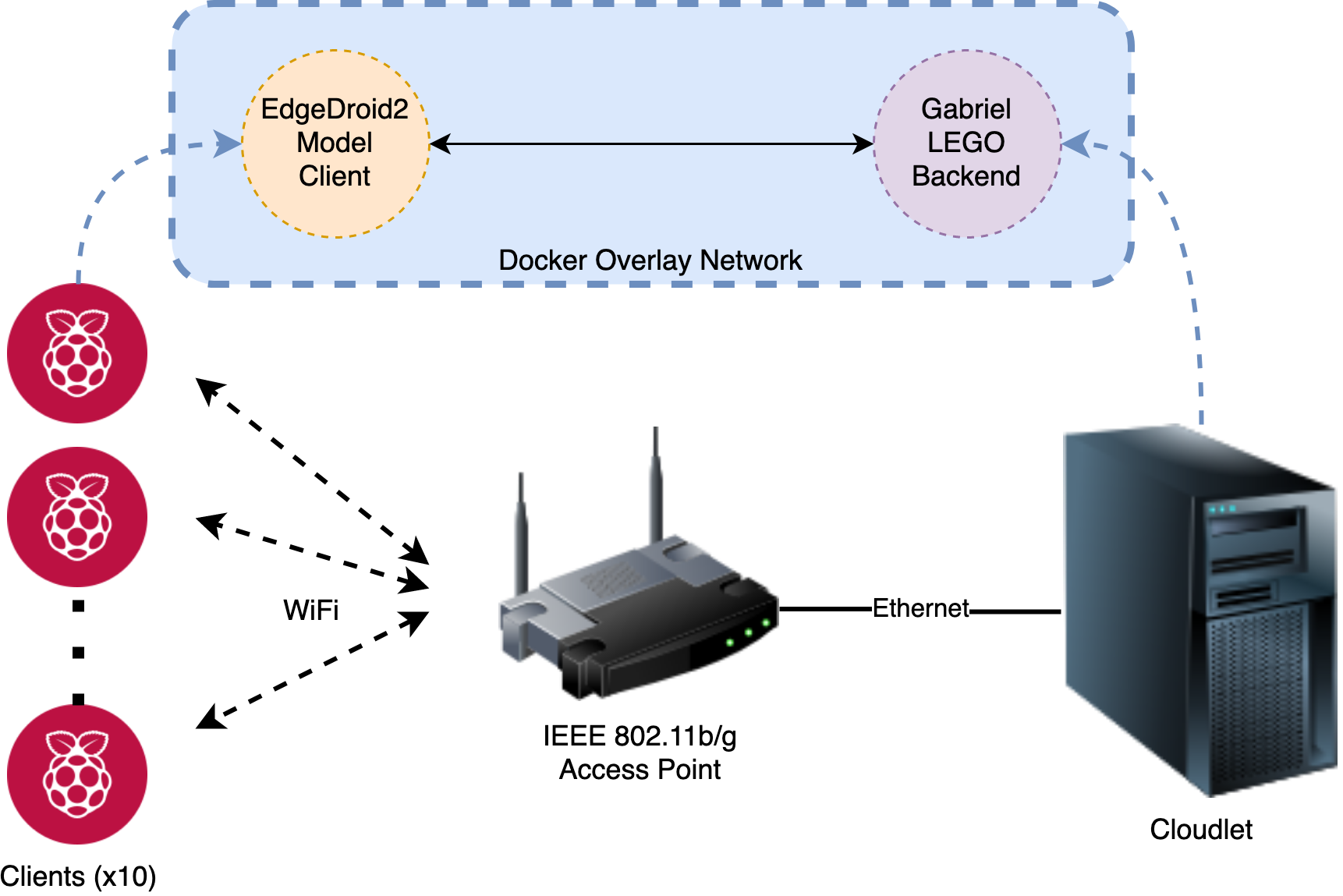}
    \caption{%
        Experimental setup used to study the implications of the realistic models of human behavior for \ac{WCA}.
        We deploy containerized instances of the client-server loop running the models on a testbed consisting of \num{10} Raspberry Pi clients connected to a cloudlet over a \acs*{COTS} \acs{IEEE} \num{802.11}b/g access point.
    }\label{fig:expsetup}
\end{figure}

Next we study the effects of first- versus second-order models in a more realistic setting.
The models, reference and realistic, are deployed on the Raspberry Pi clients of the testbed depicted in \cref{fig:expsetup}.
For this, the timing models and frame generator are integrated into a custom Python3 client for the Gabriel \ac{WCA} platform~\cite{Chen2018application}, which are then paired with real instances of Gabriel deployed on the cloudlet.
Clients and cloudlet communicate over an \acs{IEEE} \num{802.11}b/g wireless network.
Our choice of wireless standard is simply motivated by a desire to amplify the potential effects of network congestion.

\begin{figure}
    \centering
    \begin{subfigure}[]{\columnwidth}
        \centering
        \includegraphics[width=\textwidth]{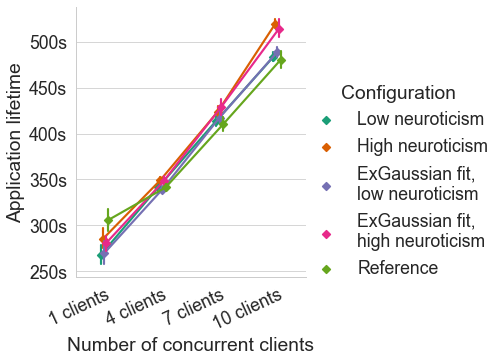}
        \caption{%
            Mean application lifetimes per testbed configuration.
            Note that due to the low number of samples, means have been calculated using the geometric instead of arithmetic average. 
            Error bars indicate \SI{95}{\percent} \acp{CI}, calculated using bootstrapping.
        }
    \end{subfigure}
    \par\bigskip
    \begin{subtable}{\columnwidth}
        \centering
        \begin{tabular}{lrrrr}
            \toprule
            \# clients & 1 & 4 & 7 & 10 \\
            \ac{RTT} & \SI{0.42}{\second} & \SI{1.12}{\second} & \SI{1.92}{\second} & \SI{2.68}{\second} \\
            \bottomrule
        \end{tabular}
        \caption{Mean measured \aclp{RTT} for each testbed configuration.}
    \end{subtable}
    \par\bigskip
    \begin{subfigure}[]{\columnwidth}
        \centering
        \includegraphics[width=\textwidth]{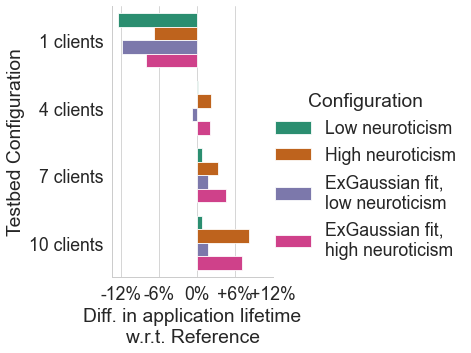}
        \caption{%
            Percentage difference in mean application lifetimes with respect to the reference model.
            Confidence intervals have been omitted due to the low number of samples and the use of the geometric mean.
        }
    \end{subfigure}
    \caption{Application lifetimes in the realistic scenarios.}\label{fig:testbed_lifetimes}
\end{figure}

We deploy configurations running \num{45}-step versions of the LEGO task described in~\cite{olguinmunoz:impact2021}.
The testbed configurations include setups with \numlist{1;4;7;10} clients.
Due to having limited time, each combination of testbed and timing model configuration is only repeated \num{10} times.
The results are presented in \cref{fig:testbed_lifetimes}.
Owing to the low number of samples, specifically to minimize the effects of potential outliers, we opt here for the geometric rather than arithmetic mean to represent our results.
The results are nonetheless clear, and follow the same pattern as the previously discussed results under ideal, controlled circumstances.
With just a single client and mean \ac{RTT} of around \SI{400}{\milli\second}, all parameterizations of the realistic model achieved an average task duration \SIrange{6}{12}{\percent} shorter than the reference.
At \num{10} clients, the results mimic those at higher \acp{TTF} in the ideal controlled setup, with high neuroticism parameterizations reaching \textasciitilde\SI{8}{\percent} longer application lifetimes.

These results highlight the importance of accurate execution time modeling when studying \ac{WCA} application lifetimes.
Not only do we see considerable differences in lifetimes at relatively moderate levels of system congestion, but the sign of these differences depends on the load placed on the system.
Imagine thus a system designer studying resource consumption optimization in a \ac{WCA}.
If they were to employ the reference model for their study on an unimpaired system, it could lead them to significantly underestimate the potential for optimization of resource consumption, leaving performance (and, potentially, monetary) gains on the table.
On the other hand, under heavy load, they would instead underestimate system resource occupation, again leading to performance losses.


\section{Implications for the optimization of resource consumption and responsiveness trade-offs in \ac{WCA}}\label{sec:implications:optimization}

As with any application intended for mass deployment on multi-tenant systems such as the edge, \ac{WCA} applications will have to be configured and designed to minimize resource consumption at runtime.
At the same time, imposing too stringent requirements on resource consumption can adversely affect application responsiveness, which in turn has the potential to drastically reduce quality of experience for users.
Optimizing the deployment of these applications thus inherently involves achieving a balance between the reduction of resource consumption while maintaining an acceptable level of system responsiveness.

A question then arises as to what extent accurate modeling of human behavior plays a role in the optimization of \ac{WCA} applications and systems.
We have already seen from the results in \cref{sec:implications:footprint} that changes in responsiveness can have drastic effects on human execution times and subsequently on overall application lifetimes.
This points towards parameters with dependencies on step timings as prime candidates for evaluating the effects of human behavior on their potential for optimization of different dimensions of \ac{WCA}.
One such parameter of particular interest to us is sampling strategy, which intrinsically hinges on a stochastic understanding of human timings in a step.
In the following section we will study the potential advantages of using the realistic model presented in \cref{sec:model} to estimate the execution time distributions of steps to adapt the sampling strategy at runtime.
We will first introduce a generic optimization framework which produces an aperiodic sampling strategy that optimizes for arbitrary responsiveness-resource consumption trade-offs in \ac{WCA}.
We will then employ this framework for the minimization of two different metrics: number of samples and total energy consumption per step.
In order to illustrate the advantages of our framework, we will compare our results with a state-of-the-art baseline which does not dynamically adapt.

\subsection{Optimization framework}\label{ssec:optframework}

Our optimization approach focuses on the individual steps which comprise a complete \ac{WCA} task.
Thus, we begin by assuming a given execution time distribution \( \mathcal{T} \), and take the modeling and partial solution approach from recents works on energy efficient sampling in edge-based feedback systems.
In \textcite{Moothedath2021EnergyOptimal,Moothedath2022EnergyEfficient}, the authors model the energy in terms of the expected number of samples $\mathbb{E}[\mathcal{S}]$ and the expected wait time $\mathbb{E}[\mathcal{W}]$ experienced by the user.
The authors then find the optimum periodic sampling interval that minimizes this energy, or equivalently the non-constant parts of this energy termed as \textit{Energy Penalty}.

With a value much smaller than the execution time of the event, the \ac{RTT} of the final sample is included in this constant part, which is why $\mathcal{W}$ is computed without it.
Next, in~\cite{Moothedath2022Aperiodic}, the author retains the model and removes the constraint of periodicity to find the optimum aperiodic sampling instants $\{t_n,\,n=1,2,\dots\}$ that minimize the same energy penalty.

In this work, we use the modeling from~\cite{Moothedath2022Aperiodic} to find the optimum \emph{aperiodic} sampling interval.
However, instead of using their two-step approach to find the solution which includes a recursive solution followed by a bisection algorithm, we develop a novel, approximate, but easier solution to finding the set of the optimum aperiodic sampling intervals.
Furthermore, instead of directly optimizing for energy, we start by noting that, in general, any objective metric in these applications which relates to sampling and the responsiveness of the system will present itself
as a linear combination between $\mathbb{E}[\mathcal{S}]$ and $\mathbb{E}[\mathcal{W}]$ plus terms independent of the number of samples or wait time.
Let $\mathcal{E}$ correspond to such an objective metric.
Thus, 
\begin{alignat}{1}
    \Rightarrow\mathcal{E}&=\alpha\mathbb{E}[\mathcal{S}]+\beta\mathbb{E}[\mathcal{W}]+C\;\label{eq:epsilon_terminal}
\end{alignat}

Here, $\alpha, \beta$ and $C$ are constants responsible for modifying the objective function from one metric to another.
For instance, in the modeling used by~\cite{Moothedath2021EnergyOptimal,Moothedath2022EnergyEfficient,Moothedath2022Aperiodic}, the chosen characterization results in a metric which is equal to the energy penalty.
Our solution which minimizes \cref{eq:epsilon_terminal} is provided in \cref{tn_approx_rayleigh}.
It assumes a $\mathcal{T}$ distributed according to a Rayleigh distribution with parameter $\sigma$.
The complete mathematical derivation that leads to this formula is detailed in \cref{appx1}.

\begin{alignat}{1}
t_n&=\Big(3\sigma\!\sqrt{\tfrac{\alpha}{2\beta}}\Big)^{\frac{2}{3}}n^{\frac{2}{3}}\label{tn_approx_rayleigh}
\end{alignat}

\subsection{Optimizing for mean number of samples per step}\label{ssec:optimization:samples}

We first look at the application of \cref{eq:epsilon_terminal} for the optimization of number of captured samples per step.
We start with this metric as its implications for resource consumption and responsiveness are straightforward to understand.
On the one hand, higher sampling rates directly lead to perceived increased system responsiveness, as smaller sampling intervals translate into smaller maximum wait times.
On the other, the relationship between number of samples captured and sent and network congestion is exponential, and too high sampling rates quickly lead to bottlenecks on the network, particularly in multi-tenant environments.
Additionally, the energy cost of capturing a sample on a \ac{WCA} client device is often much higher than remaining in an idle state, and thus excessive sampling leads to drastically increased energy consumption.
Optimizing the number of samples captured per step can thus be a straightforward way of reducing resource consumption and contention in \ac{WCA} applications.

However, an unconstrained optimization of the number of samples is trivial and meaningless as the solution points to a single sample at $t\!\rightarrow\!\infty$, which also takes the wait time to infinity.
Thus, we look at the constrained optimization of the expected number of samples with an upper bound $w_0$ for the expected wait.
That is, $\mathbb{E}[\mathcal{W}]\!\leq\!w_0$.
We show in \cref{appx1} that we can find appropriate $\alpha$ and $\beta$ for this problem by satisfying the condition
\begin{alignat}{1}\label{eq:optimization:sampling}
\frac{\alpha}{\beta}=\frac{2\sqrt{2}\,w_0^2}{(\mathlarger{\Gamma}(\tfrac{3}{4}))^2\,\sigma}\approx1.9\frac{w_0^2}{\sigma}
\end{alignat}
where $\mathlarger{\Gamma}(x)$ is the Gamma function.

\medskip

We introduce here a reference scheme to which we will compare our approach.
In~\cite{Wang2019Towards}, \citeauthor{Wang2019Towards} introduce an adaptive sampling scheme for \ac{WCA} intended to reduce the number of samples processed per step while still meeting application responsiveness bounds.
At every sampling instant \( t \), the scheme adapts the sampling rate \( R(t) \) of the system according to the estimated likelihood of the user having finished the step,
following the formula 

\begin{equation}
    R(t) = R_\text{min} + \varphi\left( R_\text{max} - R_\text{min} \right) * CDF(t)
\end{equation}

\( R_\text{max} \) and \( R_\text{min} \) correspond to the maximum and minimum sampling rates of the system, respectively.
\( R_\text{max} \) can directly be assumed to correspond to \( 1 / \text{\ac{RTT}}_\mu \), where \( \text{\ac{RTT}}_\mu \) corresponds to the mean \ac{RTT} of the system.
\( R_\text{min} \) needs to either be calculated according to the latency bounds of the system or specified manually.
\( \varphi \) corresponds to a scaling factor and \( t \) to the time of the current sampling instant with respect to the start of the step.
Finally, \( CDF \) corresponds to the \ac{CDF} of a distribution describing the execution times for the current step; \citeauthor{Wang2019Towards} used a single static Gaussian distribution for all steps in their work.

\medskip

In the following, we will show the effects of our optimization approach combined with our timing models compared to the state-of-the-art approach~\cite{Wang2019Towards}.
For this we implement these sampling schemes in Python and run a number of simulations with them.
The first scheme uses our approach, \cref{eq:epsilon_terminal,eq:optimization:sampling}, to determine the optimum sampling instants and the values for \( \alpha \) and \( \beta \) at each step.
It includes an embedded timing model (without any distribution fitting) to provide updated estimates of the mean execution time \( \mu \) and \( \sigma \) at every step as well, allowing it to adapt to the state of the system.
We will refer to this scheme as the \emph{sample-count-optimized aperiodic} sampling scheme.

We implement \citeauthor{Wang2019Towards}'s original design using a Gaussian distribution fitted to all the execution times collected for~\cite{olguinmunoz:impact2021} for \ac{CDF} calculation.
This scheme does not include an embedded timing model, and uses the same \ac{CDF} for every step.

Finally, we also implement two reference sampling schemes representing best- and worst-case extremes.
The first of these corresponds to the offline optimum which uses an embedded \emph{oracle} to perfectly predict the execution time of each step.
Such an ideal scheme is thus able to always sample exactly once per step, with a constant wait time of zero.
The second reference scheme corresponds to one which \emph{greedily} samples as much as possible.
This represents a completely unoptimized design with no considerations for resource-consumption trade-offs; it simply attempts to maximize the number of captured samples per step.
This is an interesting approach to include as it corresponds to the sampling strategy used in most existing \ac{WCA} prototypes.

\begin{table}
    \centering
    \caption{Experimental parameters}\label{tab:params}
    \begin{tabular}{lrl}
        \toprule
        Parameter & Value & Clarification \\
        \midrule
        \# of steps & \num{100} & \\
        Repetitions & \num{100} & \\
        \acp{RTT} & \( \left\{ 0.3, 0.6,\ldots,4.2 \right\} \) & \\
        \( \tau_\text{p} \) & \SI{250}{\milli\second} & Processing delay \\
        \( \tau_\text{c} \) & \( \text{\ac{RTT}} - \tau_\text{p} \) & Communication delay \\
        \( w_0 \) & \SI{1.0}{\second} & \\
        \( R_\text{min} \) & \SI{0.5}{\hertz} & Derived as \( R_\text{min} = {(2 w_0)}^{-1} \) \\
        \( \varphi \) & \num{1.5} & Scaling factor, \textcite{Wang2019Towards}\\
        \( P_0 \) & \SI{15}{\milli\watt} & Idle power \\
        \( P_\text{c} \) & \SI{45}{\milli\watt} & Communication power \\
        \( \alpha_\text{samples} \) & \( 1.9 \sigma^{-1} \) & For sample-count optimization. \\
        \( \beta_\text{samples} \) & \num{1.0} & For sample-count optimization. \\
        \( \alpha_\text{energy} \) & \( \tau_\text{c}(P_\text{c} - P_0) \) & For energy optimization. \\
        \( \beta_\text{energy} \) & \( P_0 \) & For energy optimization. \\
        \bottomrule
    \end{tabular}
\end{table}

We proceed to set up an experiment where these sampling approaches are deployed on identical, simulated, tasks with varying constant \acp{RTT}.
Experimental parameters are summarized in \cref{tab:params}.
We set the \( w_0 \) and \( \beta \) factors of our sample-count-optimized scheme to \SI{1.0}{\second} and \num{1.0}, respectively, for mathematical simplicity, and derive \( \alpha = 1.9 \sigma^{-1} \).
Next, we derive \( R_\text{min} \) for \textcite{Wang2019Towards}'s scheme from \( w_0 \).
As discussed in \cref{sec:background}, we assume that wait times are uniformly distributed between \SI{0}{\second} and sampling interval of each step.
A maximum expected wait time \( w_0 = 1.0\,\si{\second} \) thus translates into a maximum expected sampling interval of \SI{2.0}{\second}, yielding a minimum sampling rate \( R_\text{min} = {2.0\,\si{\second}}^{-1} = 0.5\,\si{\hertz} \).
It should also be noted that in both aperiodic schemes, our sampling-count-optimized approach and the \ac{CDF}-based approach, there exists the possibility for sampling instants to be \emph{missed} due to the actual \ac{RTT} of the system being higher than the parameterization of the schemes.
In these cases, both schemes will degrade into greedy sampling.

The execution times for each step are generated by a timing model without any distribution fitting on the data.
For each combination of sampling scheme configuration, \ac{RTT}, and execution time model neuroticism (low or high), we run \num{100} repetitions of the task for good statistical significance.
Note that the embedded timing model in our sampling-count-optimized aperiodic sampling scheme is always parameterized with a neuroticism matching the neuroticism of the external execution time model.

\begin{figure*}
    \centering
    \begin{subfigure}[t]{\textwidth}
        \centering
        \includegraphics[width=\textwidth]{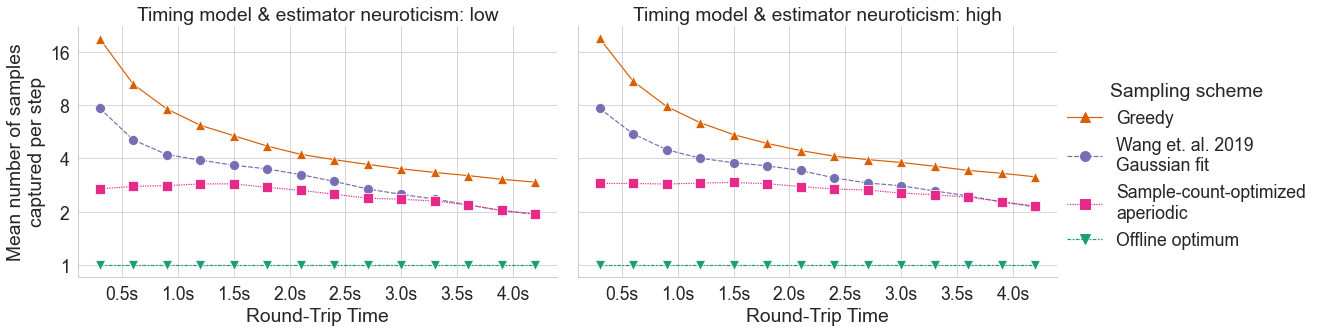}
        \caption{%
            Round-trip time versus mean number of captured samples per step, averaged over all \num{100} repetitions of the experiment.
            Note the logarithmic scale on the vertical axis.
            Error bars indicate \SI{95}{\percent} \acp{CI}.
        }
    \end{subfigure}\\
    \medskip
    \begin{subfigure}[t]{\textwidth}
        \centering
        \includegraphics[width=\textwidth]{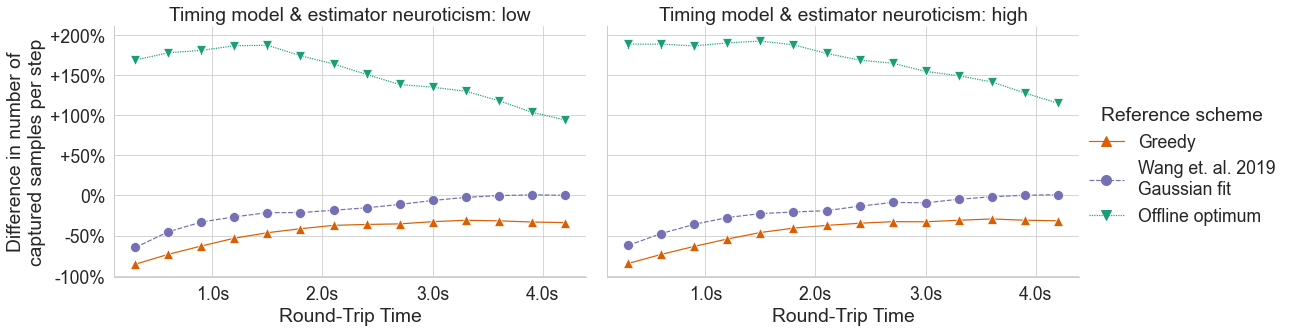}
        \caption{%
            Percentage difference in mean number of captured samples per step by the sample-count-optimized approach with respect to the three reference schemes.
            In other words, curves represent the relative performance of the sample-count-optimized scheme when using the corresponding reference scheme as baseline.
            Error bars indicate \SI{95}{\percent} \acp{CI}, calculated using a two-sided t-test.
        }
    \end{subfigure}
    \caption{%
        Summary of results for experiment comparing the sample-count-optimized aperiodic sampling scheme to the reference schemes and \textcite{Wang2019Towards}'s \ac{CDF}-based approach.
    }\label{fig:optimization:samples}
\end{figure*}

The results of this investigation are presented in \cref{fig:optimization:samples}, and clearly show the advantages of using the sample-count-optimized scheme over the current state-of-the-art.
The performance of \textcite{Wang2019Towards}'s approach appears to degrade with lower \acp{RTT}, exponentially oversampling as latency tends to zero and the maximum sampling rate of the system tends to infinity.
On the other hand our sampling scheme consistently matches or beats the state-of-the-art while maintaining a relatively constant behavior with respect to \acp{RTT}.
As mentioned above, the future feasibility and mass adoption of \ac{WCA} depends on these applications not hogging the available resources.
Our work advances this goal by being consistently more efficient than existing alternatives at minimizing the number of samples per step, and thus reducing network and processing load.

It should be noted that although both schemes seem to tend towards two samples per step as \acp{RTT} increase, this is simply an artifact of our experimental setup.
As \acp{RTT} increase above the expected wait time, the probability of the sum of the first sampling interval and the \ac{RTT} being larger than the execution time of the step tends towards \num{1.0}.
This leads to these sampling schemes consistently sampling only twice each step: a first sample which is taken before the execution time of the step, and a second one \ac{RTT} seconds later, after the execution time has been reached. 

\subsection{Optimizing for energy consumption}\label{ssec:optimization:energy}

Next we will explore the implications of combining our timing models with \cref{eq:epsilon_terminal} when directly optimizing for energy consumption in \ac{WCA}.
Although, as mentioned above, minimizing the number of samples captured during a step can \emph{potentially} translate into a reduction in the energy consumption, this is not a given.
Energy consumption depends on multiple other factors other than number of samples captured, such as idle versus communication power and delays, and thus cannot be optimized by simply minimizing the number of samples taken.

In the following, we thus take the general solution, \cref{eq:epsilon_terminal}, and find the appropriate \( \alpha \) and \( \beta \) to minimize the energy consumed per step, i.e.\ \( \mathcal{E}=E \).
We directly take the modeling from~\cite{Moothedath2022Aperiodic} with a necessary modification in the  assumption of one-way communication in all but the final sample.
With feedback given even to the discarded samples in our model and communication delay defined as the total delay in either direction, we have, 
\begin{alignat}{2}
    \mathrm{E}=&\;\mathcal{S}\tau_cP_c+(\mathcal{T}+\mathcal{W}+\tau_\mathrm{p}+\tau_\mathrm{c}-\mathcal{S}\tau_c)P_0\nonumber\\
    =&\;\tau_{\text{c}}(P_{\text{c}} -P_0)\mathcal{S}+\mathcal{W}P_0+(\mathcal{T}+\tau_{\text{p}} +\tau_{\text{c}}) P_0\nonumber\\
&\Rightarrow \alpha=\tau_{\text{c}}(P_{\text{c}} -P_0),\text{ and }\beta=P_0 \label{eq:optimization:energy}
\end{alignat}
\( \tau_\text{p} \) and \( \tau_\text{c} \) correspond to the processing and two-way communication delay for each sample.
\( P_\text{c} \) and \( P_0 \) correspond to the communication and idle power, respectively, of the \ac{WCA} client device.

We proceed to repeat the experiment detailed in \cref{ssec:optimization:samples}, replacing the sample-count-optimized sampling scheme with a new implementation instead minimizing energy, using \cref{eq:optimization:energy} for the calculation of \( \alpha \) and \( \beta \).
Once again, we embed a timing model into the sampling scheme to provide updated estimates of the mean execution time \( \mu \) at each step.
We refer to this scheme as the \emph{energy-optimized aperiodic} sampling scheme.
For the power constants, we reuse the values estimated by the authors in~\cite{Moothedath2022EnergyEfficient}, \( P_\text{c} = 45\,\si{\milli\watt} \) and \( P_\text{0} = 15\,\si{\milli\watt} \).
On the other hand, for the timing variables we define a constant processing delay \( \tau_\text{p} = 250\,\si{\milli\second} \) across all configurations and repetitions of the experiments; given then a constant \ac{RTT} for the task, we set \( \tau_\text{c} = \text{\ac{RTT}} - \tau_\text{p} \).

\begin{figure*}
    \centering
    \begin{subfigure}[t]{\textwidth}
        \centering
        \includegraphics[width=\textwidth]{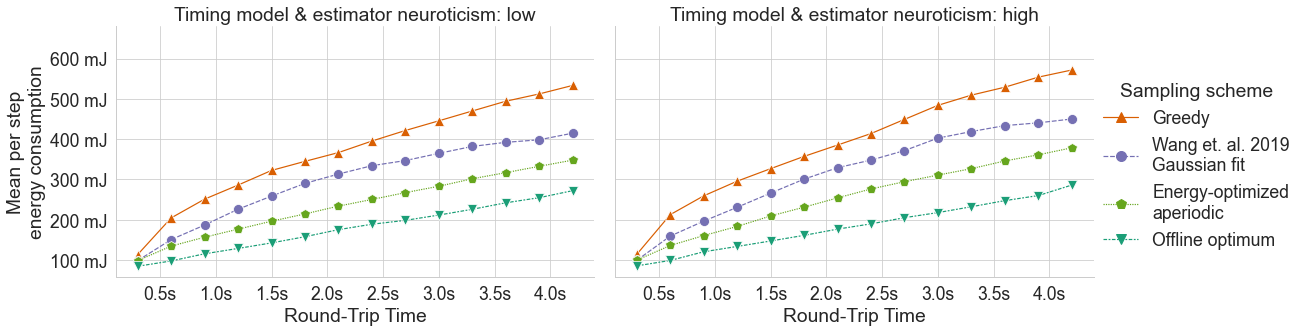}
        \caption{%
            Round-trip time versus mean per step energy consumption, averaged over all \num{100} repetitions of the experiment.
            Error bars indicate \SI{95}{\percent} \acp{CI}.
        }
    \end{subfigure}\\
    \medskip
    \begin{subfigure}[t]{\textwidth}
        \centering
        \includegraphics[width=\textwidth]{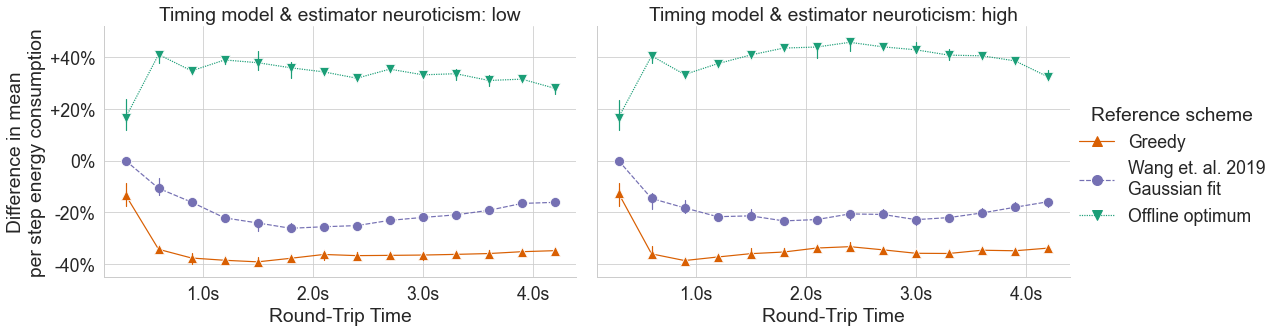}
        \caption{%
            Percentage difference in mean per step energy consumption by the energy-optimized sampling scheme with respect to the three reference schemes.
            Error bars indicate \SI{95}{\percent} \acp{CI}, calculated using a two-sided t-test.
        }
    \end{subfigure}
    \caption{%
        Summary of results for experiment comparing the energy-optimized aperiodic sampling scheme to the reference schemes and \textcite{Wang2019Towards}'s \ac{CDF}-based approach.
    }\label{fig:optimization:energy}
\end{figure*}

Again, we include the reference greedy and offline optimum schemes, as well as \textcite{Wang2019Towards}'s approach, and repeat the experiment \num{100} times for each combination of sampling scheme, neuroticism, and \ac{RTT}.

The results of these experiments are presented in \cref{fig:optimization:energy}, and they clearly illustrate the advantages of the integration of our timing models with an adaptive, energy-optimized sampling scheme when compared to unoptimized and state-of-the-art sampling schemes.
Our approach is consistently consumes \SI{20}{\percent} less energy than \textcite{Wang2019Towards}'s state-of-the-art, and is up to \SI{40}{\percent} more energy efficient than the greedy scheme.
Furthermore, once again the behavior of our approach is more consistent and reliable than the competition, exhibiting a flat curve of energy consumption much akin to that of the offline optimum in behavior.
\section{Conclusion}\label{sec:conclusion}

This paper addresses the difficulty of benchmarking \ac{WCA} by offering a model-based alternative to the extensive human-user studies that would otherwise be required.
It first introduces the \edgedroid{} model of human timing behavior for \ac{WCA}.
This model represents a stochastic approach to execution time modeling which builds upon prior data~\cite{olguinmunoz:impact2021}.
It further introduces a novel procedure for the generation of synthetic traces of frames in step-based \ac{WCA}, allowing for a full end-to-end emulation of a human when combined with the timing model.

The paper then explores the impact of such a realistic model on the application lifetime footprint of \ac{WCA} applications.
It shows that less realistic modeling approaches which do not take into account higher-order effects on execution time distributions, can potentially lead to substantial mis-estimations of the application footprint.

Finally, the paper delves into the potential for optimization in \ac{WCA} systems using the previously discussed timing models.
It proposes a novel stochastic optimization framework for resource consumption-system responsiveness trade-offs in \ac{WCA}, which results in an adaptive sampling strategy. 
We have shown that this framework is applicable to both the minimization of number of samples and total energy consumption per step by showcasing experimental results.
Our results show up to a \SI{50}{\percent} increase in performance with respect to state of the art when optimizing for number of samples, and up to a \SI{30}{\percent} improvement when optimizing for energy consumption, thus proving the value of such frameworks for the design of \ac{WCA} applications.

This work serves as an important, yet initial step towards realistic modeling of human behavior in \ac{WCA}, and more generally \ac{AR}.
Many directions remain open to additional exploration in this space.
For instance, our current model only targets a particular class of step-based \ac{WCA}, and extension of our methodology to other classes of these applications, or even more generally to \ac{AR} and \ac{XR}, is on our roadmap.
Our data~\cite{olguinmunoz:impact2021} also only considers young undergraduate students at a highly competitive university in the United States.
An extension of this dataset and the model towards a more representative sample of the general population would surely be a valuable endeavor.
Corresponding research is needed to determine individual difference factors that would substantially impact response to \ac{WCA} in novel populations.

In terms of the optimization framework, our current model adopts the assumption that execution times are Rayleigh-distributed.
Although this distribution fits the data reasonably well, there are other which more accurately describe the behavior of human execution times (e.g.\ the \acl{exGaussian}).
Our future efforts consider developing this framework towards these more accurate distributions.
Finally, our current approach only considers a single client, which will be far from the case in a real-world \ac{WCA} deployment.
As such, another milestone in this context could be the exploration for a potential extension towards a collaborative and/or distributed solution.

\section*{Acknowledgements}\label{sec:acks}

This work has been partially funded by the \ac{SSF} (grant number \verb|ITM17-0246| (ExPECA)), and the \ac{NSF} (grant number \verb|CNS-2106862|).
The funders had no role in study design, data collection and analysis, decision to publish, or preparation of the manuscript.
Any opinions, findings, conclusions or recommendations expressed in this material are those of the authors and do not necessarily reflect the view(s) of their employers or funding sources.
\printbibliography{}

@article{olguinmunoz:impact2021,
  doi       = {10.1371/journal.pone.0248690},
  author    = {Olguín Muñoz, Manuel AND Klatzky, Roberta AND Wang, Junjue AND Pillai, Padmanabhan AND Satyanarayanan, Mahadev AND Gross, James},
  journal   = {PLOS ONE},
  publisher = {Public Library of Science},
  title     = {Impact of delayed response on wearable cognitive assistance},
  year      = {2021},
  month     = {03},
  volume    = {16},
  url       = {https://doi.org/10.1371/journal.pone.0248690},
  pages     = {1-25},
  number    = {3}
}

@inproceedings{olguin2018scaling,
  title        = {Scaling on the Edge--A Benchmarking Suite for Human-in-the-Loop Applicationss},
  author       = {Olgu{\'\i}n Mu{\~n}oz, Manuel and Wang, Junjue and Satyanarayanan, Mahadev and Gross, James},
  booktitle    = {2018 IEEE/ACM Symposium on Edge Computing (SEC)},
  pages        = {323--325},
  year         = {2018},
  organization = {IEEE}
}

@inproceedings{olguin2019edgedroid,
  title     = {EdgeDroid: An experimental approach to benchmarking human-in-the-loop applications},
  author    = {Olgu{\'\i}n Mu{\~n}oz, Manuel Osvaldo J and Wang, Junjue and Satyanarayanan, Mahadev and Gross, James},
  booktitle = {Proceedings of the 20th International Workshop on Mobile Computing Systems and Applications},
  pages     = {93--98},
  year      = {2019}
}

@article{oliver:bfi1999,
  title   = {The Big Five trait taxonomy: History, measurement, and theoretical perspectives},
  author  = {John, Oliver P and Srivastava, Sanjay and others},
  journal = {Handbook of personality: Theory and research},
  volume  = {2},
  number  = {1999},
  pages   = {102--138},
  year    = {1999}
}

@inproceedings{Chen2015LEGO,
  author    = {Chen, Zhuo and Jiang, Lu and Hu, Wenlu and Ha, Kiryong and Amos, Brandon and Pillai, Padmanabhan and Hauptmann, Alex and Satyanarayanan, Mahadev},
  title     = {Early Implementation Experience with Wearable Cognitive Assistance Applications},
  year      = {2015},
  isbn      = {9781450335003},
  publisher = {Association for Computing Machinery},
  address   = {New York, NY, USA},
  url       = {https://doi.org/10.1145/2753509.2753517},
  doi       = {10.1145/2753509.2753517},
  booktitle = {Proceedings of the 2015 Workshop on Wearable Systems and Applications},
  pages     = {33-38},
  numpages  = {6},
  location  = {Florence, Italy},
  series    = {WearSys '15}
}

@inproceedings{Ha2014towards,
  author = {Ha, Kiryong and Chen, Zhuo and Hu, Wenlu and Richter, Wolfgang and Pillai, Padmanabhan and Satyanarayanan, Mahadev},
  title = {Towards Wearable Cognitive Assistance},
  year = {2014},
  isbn = {9781450327930},
  publisher = {Association for Computing Machinery},
  address = {New York, NY, USA},
  url = {https://doi.org/10.1145/2594368.2594383},
  doi = {10.1145/2594368.2594383},
  booktitle = {Proceedings of the 12th Annual International Conference on Mobile Systems, Applications, and Services},
  pages = {68-81},
  numpages = {14},
  location = {Bretton Woods, New Hampshire, USA},
  series = {MobiSys '14}
}

@article{Satya2019augmenting,
  title     = {Augmenting cognition through edge computing},
  author    = {Satyanarayanan, Mahadev and Davies, Nigel},
  journal   = {Computer},
  volume    = {52},
  number    = {7},
  pages     = {37--46},
  year      = {2019},
  publisher = {IEEE},
  doi={10.1109/MC.2019.2911878}
}

@phdthesis{Chen2018application,
  title  = {An application platform for wearable cognitive assistance},
  author = {Chen, Zhuo},
  year   = {2018},
  school = {Carnegie Mellon University},
  url = {https://elijah.cs.cmu.edu/DOCS/CMU-CS-18-104.pdf}
}

@inproceedings{Wang2019Towards,
  title     = {Towards scalable edge-native applications},
  author    = {Wang, Junjue and Feng, Ziqiang and George, Shilpa and Iyengar, Roger and Pillai, Padmanabhan and Satyanarayanan, Mahadev},
  booktitle = {Proceedings of the 4th ACM/IEEE Symposium on Edge Computing},
  pages     = {152--165},
  year      = {2019}
}

@article{Palmer2011shapes,
  title     = {What are the shapes of response time distributions in visual search?},
  author    = {Palmer, Evan M and Horowitz, Todd S and Torralba, Antonio and Wolfe, Jeremy M},
  journal   = {Journal of experimental psychology: human perception and performance},
  volume    = {37},
  number    = {1},
  pages     = {58},
  year      = {2011},
  publisher = {American Psychological Association}
}

@article{Rohrer1994analysis,
  title     = {An analysis of latency and interresponse time in free recall},
  author    = {Rohrer, Doug and Wixted, John T},
  journal   = {Memory \& Cognition},
  volume    = {22},
  number    = {5},
  pages     = {511--524},
  year      = {1994},
  publisher = {Springer}
}

@article{Marmolejo2022generalised,
  title     = {Generalised exponential-Gaussian distribution: a method for neural reaction time analysis},
  author    = {Marmolejo-Ramos, Fernando and Barrera-Causil, Carlos and Kuang, Shenbing and Fazlali, Zeinab and Wegener, Detlef and Kneib, Thomas and De Bastiani, Fernanda and Martinez-Florez, Guillermo},
  journal   = {Cognitive Neurodynamics},
  pages     = {1--17},
  year      = {2022},
  publisher = {Springer}
}

@ARTICLE{Moothedath2022EnergyEfficient,
  author={Moothedath, Vishnu Narayanan and Champati, Jaya Prakash Varma and Gross, James},
  journal={IEEE Transactions on Mobile Computing},
  title={Energy Efficient Sampling Policies for Edge Computing Feedback Systems},   year={2022},
  volume={},
  number={},
  pages={1-1},
  doi={10.1109/TMC.2022.3165852}
}

@INPROCEEDINGS{Moothedath2021EnergyOptimal,
  author={Moothedath, Vishnu Narayanan and Champati, Jaya Prakash and Gross, James},  
  booktitle={IEEE International Conference on Communications Workshops (ICC Workshops)},
  title={Energy-Optimal Sampling of Edge-Based Feedback Systems},
  year={2021},
  volume={},
  number={},
  pages={1-6},
  doi={10.1109/ICCWorkshops50388.2021.9473894}
}

@article{Satoshi1992Optimal,
author="Satoshi, Fukumoto and Naoto, Kaio and Shunji, Osaki",
title="Optimal Checkpointing Policies Using the Checkpointing Density",
journal="Journal of Information Processing",
ISSN="1882-6652",
year="1992",
month="03",
volume="15",
number="1",
pages="97-92"
}

@article{Bellman1954Dynamic,
  title={Dynamic programming and a new formalism in the calculus of variations},
  author={Bellman, Richard},
  journal={Proceedings of the National Academy of Sciences of the United States of America},
  volume={40},
  number={4},
  pages={231},
  year={1954},
  publisher={National Academy of Sciences}
}

@inbook{Arfken2015Calculus,
author = {Arfken, George and Weber, Hans and Harris, Frank},
year = {2013},
month = {12},
pages = {1081-1124},
title = {Calculus of Variations},
isbn = {9780123846549},
}

@INPROCEEDINGS{Moothedath2022Aperiodic,
  author={Moothedath, Vishnu Narayanan},
  booktitle={IEEE/ACM Symposium on Edge Computing (SEC)},
  title={Energy-Optimal Sampling for Edge Computing Feedback Systems: Aperiodic Case (To be published)},
  year={2022},
  volume={},
  number={},
  pages={403-408},
  doi={}
}

@inproceedings{Chen2017Empirical,
  title     = {An empirical study of latency in an emerging class of edge computing applications for wearable cognitive assistance},
  author    = {Chen, Zhuo and Hu, Wenlu and Wang, Junjue and Zhao, Siyan and Amos, Brandon and Wu, Guanhang and Ha, Kiryong and Elgazzar, Khalid and Pillai, Padmanabhan and Klatzky, Roberta and others},
  booktitle = {Proceedings of the Second ACM/IEEE Symposium on Edge Computing},
  pages     = {1--14},
  year      = {2017}
}

@article{Funk2015Cognitive,
  title     = {Cognitive assistance in the workplace},
  author    = {Funk, Markus and Schmidt, Albrecht},
  journal   = {IEEE Pervasive Computing},
  volume    = {14},
  number    = {3},
  pages     = {53--55},
  year      = {2015},
  publisher = {IEEE}
}

@article{Wang2022Comprehensive,
  title     = {A comprehensive review of augmented reality-based instruction in manual assembly, training and repair},
  author    = {Wang, Zhuo and Bai, Xiaoliang and Zhang, Shusheng and Billinghurst, Mark and He, Weiping and Wang, Peng and Lan, Weiqi and Min, Haitao and Chen, Yu},
  journal   = {Robotics and Computer-Integrated Manufacturing},
  volume    = {78},
  pages     = {102407},
  year      = {2022},
  publisher = {Elsevier}
}

@article{witmer1998measuring,
  title     = {Measuring presence in virtual environments: A presence questionnaire},
  author    = {Witmer, Bob G and Singer, Michael J},
  journal   = {Presence},
  volume    = {7},
  number    = {3},
  pages     = {225--240},
  year      = {1998},
  publisher = {MIT Press}
}

@article{hirsh2008delay,
  title     = {Delay discounting: Interactions between personality and cognitive ability},
  author    = {Hirsh, Jacob B and Morisano, Dominique and Peterson, Jordan B},
  journal   = {Journal of research in personality},
  volume    = {42},
  number    = {6},
  pages     = {1646--1650},
  year      = {2008},
  publisher = {Elsevier}
}

@article{massey1951kolmogorov,
  title     = {The Kolmogorov-Smirnov test for goodness of fit},
  author    = {Massey Jr, Frank J},
  journal   = {Journal of the American statistical Association},
  volume    = {46},
  number    = {253},
  pages     = {68--78},
  year      = {1951},
  publisher = {Taylor \& Francis}
}

@article{lecci2021open,
  title     = {An Open Framework for Analyzing and Modeling {XR} Network Traffic},
  author    = {Lecci, Mattia and Drago, Matteo and Zanella, Andrea and Zorzi, Michele},
  journal   = {IEEE Access},
  volume    = {9},
  pages     = {129782--129795},
  year      = {2021},
  publisher = {IEEE}
}

@article{belletier2021wearable,
  title     = {Wearable cognitive assistants in a factory setting: a critical review of a promising way of enhancing cognitive performance and well-being},
  author    = {Belletier, Cl{\'e}ment and Charkhabi, Morteza and Pires de Andrade Silva, Gustavo and Ametepe, Kevin and Lutz, Mathieu and Izaute, Marie},
  journal   = {Cognition, Technology \& Work},
  volume    = {23},
  number    = {1},
  pages     = {103--116},
  year      = {2021},
  publisher = {Springer}
}

@article{satyanarayanan2009case,
  title     = {The case for vm-based cloudlets in mobile computing},
  author    = {Satyanarayanan, Mahadev and Bahl, Paramvir and Caceres, Ram{\'o}n and Davies, Nigel},
  journal   = {IEEE pervasive Computing},
  volume    = {8},
  number    = {4},
  pages     = {14--23},
  year      = {2009},
  publisher = {IEEE}
}

@inproceedings{chetoui2022arbench,
  title        = {{ARBench}: Augmented Reality Benchmark For Mobile Devices},
  author       = {Chetoui, Sofiane and Shahi, Rahul and Abdelaziz, Seif and Golas, Abhinav and Hijaz, Farrukh and Reda, Sherief},
  booktitle    = {2022 IEEE International Symposium on Performance Analysis of Systems and Software (ISPASS)},
  pages        = {242--244},
  year         = {2022},
  organization = {IEEE}
}

@article{george2020openrtist,
  title     = {{OpenRTiST}: end-to-end benchmarking for edge computing},
  author    = {George, Shilpa and Eiszler, Thomas and Iyengar, Roger and Turki, Haithem and Feng, Ziqiang and Wang, Junjue and Pillai, Padmanabhan and Satyanarayanan, Mahadev},
  journal   = {IEEE Pervasive Computing},
  volume    = {19},
  number    = {4},
  pages     = {10--18},
  year      = {2020},
  publisher = {IEEE}
}

@article{jing2019neural,
  title     = {Neural style transfer: A review},
  author    = {Jing, Yongcheng and Yang, Yezhou and Feng, Zunlei and Ye, Jingwen and Yu, Yizhou and Song, Mingli},
  journal   = {IEEE transactions on visualization and computer graphics},
  volume    = {26},
  number    = {11},
  pages     = {3365--3385},
  year      = {2019},
  publisher = {IEEE}
}

@inproceedings{srinivasan2009performance,
  title        = {Performance characterization and optimization of mobile augmented reality on handheld platforms},
  author       = {Srinivasan, Sadagopan and Fang, Zhen and Iyer, Ravi and Zhang, Steven and Espig, Mike and Newell, Don and Cermak, Daniel and Wu, Yi and Kozintsev, Igor and Haussecker, Horst},
  booktitle    = {2009 IEEE International Symposium on Workload Characterization (IISWC)},
  pages        = {128--137},
  year         = {2009},
  organization = {IEEE}
}

@inproceedings{huang2021proactive,
  title        = {Proactive edge cloud optimization for mobile augmented reality applications},
  author       = {Huang, Zhaohui and Friderikos, Vasilis},
  booktitle    = {2021 IEEE Wireless Communications and Networking Conference (WCNC)},
  pages        = {1--6},
  year         = {2021},
  organization = {IEEE}
}

@inproceedings{cooper2010benchmarking,
  title     = {Benchmarking cloud serving systems with YCSB},
  author    = {Cooper, Brian F and Silberstein, Adam and Tam, Erwin and Ramakrishnan, Raghu and Sears, Russell},
  booktitle = {Proceedings of the 1st ACM symposium on Cloud computing},
  pages     = {143--154},
  year      = {2010}
}

@inproceedings{jia2013characterizing,
  title        = {Characterizing data analysis workloads in data centers},
  author       = {Jia, Zhen and Wang, Lei and Zhan, Jianfeng and Zhang, Lixin and Luo, Chunjie},
  booktitle    = {2013 IEEE International Symposium on Workload Characterization (IISWC)},
  pages        = {66--76},
  year         = {2013},
  organization = {IEEE}
}

@inproceedings{das2018edgebench,
  title        = {Edgebench: Benchmarking edge computing platforms},
  author       = {Das, Anirban and Patterson, Stacy and Wittie, Mike},
  booktitle    = {2018 IEEE/ACM International Conference on Utility and Cloud Computing Companion (UCC Companion)},
  pages        = {175--180},
  year         = {2018},
  organization = {IEEE}
}

@inproceedings{mcchesney2019defog,
  title     = {Defog: fog computing benchmarks},
  author    = {McChesney, Jonathan and Wang, Nan and Tanwer, Ashish and De Lara, Eyal and Varghese, Blesson},
  booktitle = {Proceedings of the 4th ACM/IEEE Symposium on Edge Computing},
  pages     = {47--58},
  year      = {2019}
}

@article{al2017energy,
  title     = {Energy-efficient resource allocation for mobile edge computing-based augmented reality applications},
  author    = {Al-Shuwaili, Ali and Simeone, Osvaldo},
  journal   = {IEEE Wireless Communications Letters},
  volume    = {6},
  number    = {3},
  pages     = {398--401},
  year      = {2017},
  publisher = {IEEE}
}
\appendix{}\label[appendixwithoutnumber]{appx1}

In this appendix, we give the mathematical derivation of the optimum aperiodic sampling interval discussed in \cref{ssec:optframework}.
We start with the general solution, where the objective function to be minimized is given by \cref{eq:epsilon_terminal}:
\begin{alignat}{1}
    \Rightarrow\mathcal{E}&=\alpha\mathbb{E}[\mathcal{S}]+\beta\mathbb{E}[\mathcal{W}]+C\nonumber
\end{alignat}

We first borrow the idea of checkpointing density from~\cite{Satoshi1992Optimal} and define an instantaneous sampling rate function $r(t)$ which is related to $\{t_n\}$ such that,
\begin{alignat}{1}
{\int_{t_{n-1}}^{t_n}}r(t)\dif t=1,\;\forall n\geq1\label{rt}
\end{alignat}
Note that, for periodic sampling, this function is a constant, equal to the sampling frequency.
In the aperiodic case, we find $r^*(t)$, the $r(t)$ that minimizes $\mathcal{E}$.
By construction, the number of samples taken up to any time instant can be computed directly by computing the area under $r(t)$.
Thus, we obtain the expected number of samples $\mathbb{E}[\mathcal{S}]$ as
\begin{alignat}{1}
\mathbb{E}[\mathcal{S}]&=\int_{t=0}^{\infty}\bigg(\!\int_{x=0}^{t}\!\!\!\!r(x)\,\mathrm{d}x\bigg)f_{\mathcal{T}}(t)\,\mathrm{d}t.\label{Es}
\end{alignat}

To find 
$\mathbb{E}[\mathcal{W}]$, we use the conditional \ac{CDF} of the execution time.
\begin{multline*}
\mathbb{P}(\mathcal{W}=t_n-\mathcal{T}\leq t\,\big\vert\,t_{n-1}<\mathcal{T}\leq t_n)\\=\dfrac{\mathbb{P}(\mathcal{T}\geq t_n-t\,,\,t_{n-1}<\mathcal{T}\leq t_n)}{\mathbb{P}(t_{n-1}<\mathcal{T}\leq t_n)}.
\end{multline*}
The numerator is degenerate when $t\!<\!0$ or $t\!>\!(t_n\!-\!t_{n-1})$.
Thus, we are only interested in $0\!\leq\!t\!\leq\! (t_n-t_{n-1})$.
Let $F_{\mathcal{T}}$, $\bar{F}_{\mathcal{T}}$ and $f_{\mathcal{T}}$ correspond to the \ac{CDF}, \ac{CCDF} and \ac{PDF} of the execution time distribution.
\begin{alignat}{1}
\!\!\!\Rightarrow\mathbb{P}(\mathcal{W}\leq t\,\big\vert\,t_{n-1}<\mathcal{T}\leq t_n)&=\dfrac{\mathbb{P}(t_n-t\leq\mathcal{T}\leq t_n)}{F_\mathcal{T}(t_n)-F_\mathcal{T}(t_{n-1})}\nonumber\\
&\approx\dfrac{F_\mathcal{T}(t_n)-F_\mathcal{T}(t_{n}-t)}{F_\mathcal{T}(t_n)-F_\mathcal{T}(t_{n-1})}\label{Apx1}
\end{alignat}
Here, \cref{Apx1} is an approximation merely for mathematical maturity due to the slackness of the first inequality in the numerator.
We expand $F_\mathcal{T}(t_{n}-t)$ and $F_\mathcal{T}(t_{n-1})$ using Taylor series. Thus we can write the \ac{CCDF} as
\begin{multline*}
    =\Big(F_\mathcal{T}(t_n)-\\\big(F_\mathcal{T}(t_n)+f_\mathcal{T}(t_n)(-t)+f'_\mathcal{T}(t_n)(-t)^2/2!+\dots\big)\Big)\\
    \div \Big(F_\mathcal{T}(t_n)-\big(F_\mathcal{T}(t_n)+f_\mathcal{T}(t_n)(t_{n-1}-t_n)\\+f'_\mathcal{T}(t_n)(t_{n-1}-t_n)^2/2!+\dots\big)\Big).
\end{multline*}
Simplifying and approximating by ignoring the higher order terms, we arrive at
\begin{alignat}{1}
\mathbb{P}(\mathcal{W}\leq t\,\big\vert\,t_{n-1}<\mathcal{T}\leq t_n)&\approx\dfrac{tf_\mathcal{T}(t_n)}{(t_{n}-t_{n-1})f_\mathcal{T}(t_n)}\label{Apx2}\\
\Rightarrow\mathbb{P}(\mathcal{W}> t\,|\,t_{n-1}<\mathcal{T}\leq t_n)&= 1-\dfrac{t}{(t_{n}-t_{n-1})}\nonumber
\end{alignat}

Next, using the above \ac{CCDF}, we find the conditional expectation of $\mathcal{W}$.
\begin{alignat*}{1}
\Rightarrow \mathbb{E}[\mathcal{W}\,|\,t_{n-1}<\mathcal{T}\leq t_n]&=\smashoperator[r]{\int_{0}^{t_n-t_{n-1}}}\Big(1-\dfrac{t}{(t_{n}-t_{n-1})}\Big)\,\mathrm{d}t\\
&=\dfrac{(t_n-t_{n-1})}{2}.
\end{alignat*}
If $r(t)$ is varying slowly between two consecutive sampling instants due to the closeness of two sampling intervals, we can approximate the sampling interval $(t_n\!-\!t_{n-1})$ as
\begin{alignat}{1}
 (t_n-t_{n-1})&\approx\dfrac{1}{r(t)},\;\forall t,n:t_{n-1}\!<\!t\!\leq\!t_n.\label{Apx3}\\
\Rightarrow \mathbb{E}[\mathcal{W}]&=\int_{0}^{\infty}\mathbb{E}[\mathcal{W}\,|\,t_{n-1}<\mathcal{T}\leq t_n]f_\mathcal{T}(t)\,\mathrm{d}t\nonumber\\
&=\int_{0}^{\infty}\dfrac{1}{2r(t)}f_\mathcal{T}(t)\,\mathrm{d}t.\label{Ew}
\end{alignat}
We can thus find the energy penalty using \cref{Es} and \cref{Ew} as
\begin{alignat*}{1}
\mathcal{E}&=\int_{0}^{\infty}\Big(\alpha\int_{x=0}^{t}r(x)\,\mathrm{d}x+\beta\dfrac{1}{2r(t)}\Big)f_{\mathcal{T}}(t)\,\mathrm{d}t.\label{epsilon_eulerForm}\\
\intertext{%
    Let $g(t)=\int_{0}^{t}r(x)\dif x$.
    Then  $g'(t)=\tfrac{\mathrm{d}}{\mathrm{d}t}g(t)=r(t)$.
    That is,
}
\mathcal{E}&=\int_{0}^{\infty}\Big(\alpha g(t)+\dfrac{\beta}{2g'(t)}\Big)f_{\mathcal{T}}(t)\,\mathrm{d}t
\end{alignat*}

As per the Euler-Lagrange equation from the calculus of variations~\cite{Bellman1954Dynamic,Arfken2015Calculus}, the extreme value of $\mathcal{E}$ is obtained at 
\begin{alignat}{1}
r^*(t)&=\sqrt{\dfrac{\beta f_\mathcal{T}(t)}{2\alpha\bar{F}_\mathcal{T}(t)}}.\\
\intertext{Thus, for a Rayleigh distributed $\mathcal{T}$ with parameter $\sigma$,}
r^*(t)&=\sqrt{\dfrac{\beta t}{2\alpha\sigma^2}}\\
\Rightarrow &\int_{t_n}^{t_{n+1}}\!\!\!\sqrt{\dfrac{\beta t}{2\alpha\sigma^2}}\,\mathrm{d}t=1,\;\forall n\geq1\nonumber\tag{from \eqref{rt}}\\
\Rightarrow &\;t_{n+1}^{\frac{3}{2}}-t_{n}^{\frac{3}{2}}=3\sigma\!\sqrt{\tfrac{\alpha}{2\beta}}\nonumber\\
\intertext{We have $t_0=0$. Substituting $n=1,2,\dots$, in order, in the above equation provides us our final result}
&\;t_n=\Big(3\sigma\!\sqrt{\tfrac{\alpha}{2\beta}}\Big)^{\frac{2}{3}}n^{\frac{2}{3}}\label{tnRayleigh}.
x\end{alignat}
Note that, due to the close similarity in their density functions, the task times can be approximated fairly equally to an \acl{exGaussian} distribution as well as a Rayleigh distribution, with the former attracting more attention from works like~\cite{Rohrer1994analysis,Palmer2011shapes,Marmolejo2022generalised}.
This is the reason why the above results are applicable in this work where we have predominantly considered \ac{exGaussian} distribution.
Furthermore, we have also verified the closeness of the results as well as the validity of the approximations made in the proofs using distribution fitting and simulations.

By making use of this general solution, we can also prove the results given in \cref{ssec:optimization:samples}, where we modify the problem and find the optimum sampling instants that minimize the expected number of samples for a given upper bound $w_0$ for the expected wait time.
First note that the general solution has constants $\alpha$ and $\beta$ corresponding to the weights given to the cost of sampling and waiting, respectively.
The optimization criteria changes from minimizing wait time to minimizing the number of samples when the ratio $\frac{\alpha}{\beta}$ goes from zero to infinity.
Since any positive real value is valid for this ratio, one can achieve any valid point $(\mathbb{E}[\mathcal{S}],\mathbb{E}[\mathcal{W}])$ via simply by varying the ratio $\frac{\alpha}{\beta}$.
\newpage%
Furthermore, since the sampling instants are aperiodic and can take any positive real values, the bound will be tight at the optimum. 
Hence, to solve for the modified optimization problem explained in \cref{ssec:optimization:samples} that minimizes $\mathbb{E}[\mathcal{S}]$ with an upper bound $w_0$ on $\mathbb{E}[\mathcal{W}]$, we equate \cref{Ew} to $w_0$, find the corresponding $\frac{\alpha}{\beta}$, and find the optimum set of sampling instants by plugging this ratio into \cref{tnRayleigh}.
\begin{alignat*}{1}
\mathbb{E}[\mathcal{W}]&=\int_{0}^{\infty}\dfrac{1}{2r(t)}f_\mathcal{T}(t)\,\mathrm{d}t.\nonumber\\
&=\int_{0}^{\infty}\frac{1}{2}\sqrt{\frac{2\alpha\sigma^2}{\beta t}}\cdot \frac{t}{\sigma^2}e^{-{t^2}/{2\sigma^2}}\mathrm{d}t\\
&=\sqrt{\frac{\alpha\sigma^2}{2\beta}}\int_{0}^{\infty}\frac{\sqrt{t}}{\sigma^2}e^{-{t^2}/{2\sigma^2}}\mathrm{d}t\\
&=\sqrt{\frac{\alpha\sigma^2}{2\beta}}\left({\frac{1}{{2}\sigma^2}}\right)^{1/4}\cdot\int_{0}^{\infty}y^{-1/4}e^{-y}\mathrm{d}y\\
&=\sqrt{\frac{\alpha\sigma^2}{2\beta}}\left({\frac{1}{{2}\sigma^2}}\right)^{1/4}\cdot\mathlarger{\Gamma}(\tfrac{3}{4}),\\
\intertext{%
    where $\mathlarger{\Gamma(x)}$ is the gamma function.
    Thus, when the upper bound $w_0$ is tight,
}
\mathbb{E}[\mathcal{W}]&=w_0=\sqrt{\frac{\alpha\sigma}{2\sqrt{2}\beta}}\mathlarger{\Gamma}(\tfrac{3}{4})\\
\Rightarrow\frac{\alpha}{\beta}&=\frac{w_0^2}{\sigma}\frac{2\sqrt{2}}{(\mathlarger{\Gamma}(\tfrac{3}{4}))^2}\\
&\approx1.9\frac{w_0^2}{\sigma}.
\end{alignat*}

\end{document}